\documentclass{article}

\usepackage{PRIMEarxiv}

\usepackage{abstract}
\usepackage[utf8]{inputenc} 
\usepackage[T1]{fontenc}   
\usepackage{hyperref}       
\usepackage{url}            
\usepackage{booktabs}       
\usepackage{amsfonts}       
\usepackage{amsmath}        
\usepackage{amsthm}         
\usepackage{nicefrac}       
\usepackage{microtype}      
\usepackage{lipsum}
\usepackage{fancyhdr}       
\usepackage{graphicx}       
\graphicspath{{media/}}     

\newtheorem{proposition}{Proposition}[section]
\newtheorem{theorem}{Theorem}[section]

\title{A QUBO Formulation for the Generalized LinkedIn Queens and Takuzu/Tango Game
 }

\author{
  Alejandro Mata Ali \\
  Instituto Tecnológico de Castilla y León (ITCL), Spain\\
  Quantum Information and Quantum Computing Group, QuantumQuipu,\\
  National University of San Marcos, Av. Germán Amézaga s/n. Ciudad Universitaria, Lima, Perú\\
  \texttt{alejandro.mata@itcl.es} \\
  \And
  Edgar Mencia \\
  Quantum Information and Quantum Computing Group, QuantumQuipu,\\
  National University of San Marcos, Av. Germán Amézaga s/n. Ciudad Universitaria, Lima, Perú\\
  Thinkcomm SRL, (1527) Asunción, Paraguay\\
  \texttt{edmenciab@gmail.com} \\
}

\begin{document}
\maketitle

\begin{abstract}
In this paper, we present a QUBO formulation designed to solve a series of generalisations of the LinkedIn queens game, a version of the N-queens problem, for the Takuzu game (or Binairo), for the most recent LinkedIn game, Tango, and for its generalizations. We adapt this formulation for several particular cases of the problem, as Tents \& Trees, by trying to optimise the number of variables and interactions, improving the possibility of applying it on quantum hardware by means of Quantum Annealing or the Quantum Approximated Optimization Algorithm (QAOA). We also present two new types of problems, the Coloured Chess Piece Problem and the Max Chess Pieces Problem, with their corresponding QUBO formulations.
\end{abstract}

\tableofcontents
\section{Introduction}
Combinatorial games have always been an inspiration for developing and perfecting new mathematical and computational problem-solving techniques, as well as for benchmarking. Combinatorial problems are known for their practical application in academic and industrial contexts, but also for their usefulness in creating entertaining games. Some of these are the Akari, the KenKen, the Takuzu or the N-queens, which led to different combinatorial games, among which is the LinkedIn Queens.

The N-queens problem \cite{N_Queens} is a problem that has attracted enormous attention since its proposal in 1848 by the chess player Max Bezzel, leading many mathematicians, including Gauss, to try to solve and generalise it. This problem can be solved efficiently, while the problem of determining how many solutions exist becomes much more complicated. Such a problem has no applications, but it is a good problem on which to test methods of solving logically constrained problems.

LinkedIn recently published its Queens game (which we will call LQueens), based on N-queens problem, but with some changes in the restrictions to be carried out and the initial conditions. In this new version, the restriction on diagonals is eased, but a restriction on regions is imposed. Following the success that this game has had among the community, LinkedIn announced its new game called Tango, a version of the Takuzu game (also named Binairo or 0h h1) for a $6\times 6$ board, which consists of placing a set of suns and moons on a board following a series of constraints. This game can be visualized as a single-player version of tic-tac-toe, and has some applications such as the modeling of quantum agents \cite{Takuzu_quantum}. This problem has been solved previously using genetic algorithms  \cite{Binary_Genetic}.

In recent years, quantum computing has gained great popularity due to its different computational capabilities compared to classical computing, making it possible to tackle certain problems more efficiently. Some of these problems are combinatorial optimisation problems, in which quantum annealing using quantum annealers \cite{Annealing_Overview,Annealing_Industry} such as those of Dwave and the Quantum Approximated Optimization Algorithm (QAOA) \cite{QAOA} using digital quantum computers stand out. In both cases, the formulation of the problem comes through a quadratic unconstrained optimisation problem (QUBO) \cite{QUBO}, which contains the interactions to be optimised between the different variables, represented by logical qubits, minimising the energy of the system to find the lowest cost combination. These methods can be adapted to cases with constraints by adding terms that increase the energy substantially if the constraints are not respected. An example can be the QUBO formulation of the Sudoku game \cite{Sudoku_QUBO}.

Recently, the N-queens problem has been solved with a quantum algorithm in a quantum simulator \cite{Quantum_N_Queens}, with two quantum algorithms based on the W-states and backtracking \cite{Quantum_Alg_Queens}, and has also been solved with several QUBO formulations \cite{Old_QUBO_N_queens, At_most_QUBO_queens}. However, these solutions are tailored and restricted to the N-queens, and not to all generalizations we may have of the problem.

Motivated by this new version of the game and the quantum computing algorithms, we present a series of generalisations of the N-queens problem and a QUBO formulation associated with each of them so that they can be solved by quantum means, such as quantum annealing or the Quantum Approximated Optimization Algorithm (QAOA). With these generalizations, we can define two new types of problems, which take advantage of the definitions and terms we created for the N-queens generalizations. Due to the popularity of Takuzu as a combinatorial problem and its connection to quantum computing, it is also interesting to propose a QUBO formulation for Takuzu, Tango and generalizations of it that we want to formulate.

\subsection{QUBO preliminaries}
For binary variables $x_i\in\{0,1\}$ we will repeatedly use the identity $x_i^2=x_i$, so every expanded penalty below is a genuine QUBO with at most linear and quadratic terms. Let
\[
    s=\sum_{i\in I} x_i
\]
and let $A>0$ be a penalty coefficient.

\begin{itemize}
    \item Exact constraint $s=k$:
    \begin{equation}
        P_{=k}(s)=A(s-k)^2.
    \end{equation}
    Since $s$ is an integer, $P_{=k}(s)\geq 0$ and it vanishes exactly when $s=k$.

    \item At most one active variable:
    \begin{equation}
        P_{\leq 1}(I)=A\sum_{\substack{i<j\\ i,j\in I}} x_i x_j.
    \end{equation}
    Each product contributes 1 exactly when two variables are simultaneously active, so the minimizers are precisely the assignments with $\sum_{i\in I}x_i\leq 1$.

    \item Exactly $q$ or $q+1$ active variables, without slack variables:
    \begin{equation}\label{eq: prelim q qplus}
        P_{q,q+1}(s)=A\left(s-q-\frac{1}{2}\right)^2.
    \end{equation}
    Because $s$ is integer, the minimizers of \eqref{eq: prelim q qplus} are exactly $s\in\{q,q+1\}$. Its minimum value is $A/4$, not $0$. If one wants zero energy on feasible assignments, one may subtract the constant $A/4$ or equivalently use
    \begin{equation}\label{eq: prelim q qplus zero}
        P'_{q,q+1}(s)=A(s-q)(s-q-1),
    \end{equation}
    which has the same minimizers and differs from \eqref{eq: prelim q qplus} only by the additive constant $-A/4$.

    \item No three equal consecutive values. For binary variables $a,b,c$ let $s=a+b+c$:
    \begin{equation}
        P_{\mathrm{triple}}(a,b,c)=A(s-1)(s-2).
    \end{equation}
    Since $s\in\{0,1,2,3\}$, this penalty is positive only for $s=0$ and $s=3$, namely for the forbidden patterns $000$ and $111$, and it is zero for the other six assignments.

    \item Equality $x=y$:
    \begin{equation}
        P_{=}(x,y)=A(x-y)^2.
    \end{equation}
    This vanishes exactly when $x=y$.

    \item Difference $x\neq y$:
    \begin{equation}
        P_{\neq}(x,y)=A(x+y-1)^2.
    \end{equation}
    For binary variables, $x+y-1=0$ if and only if $(x,y)\in\{(0,1),(1,0)\}$.

    \item Conflict graph. If $E$ is a set of forbidden simultaneous activations,
    \begin{equation}
        P_E(x)=A\sum_{\{u,v\}\in E} x_u x_v.
    \end{equation}
    Each conflicting edge contributes exactly when both endpoints are active, so the minimizers are the independent sets of the conflict graph.
\end{itemize}

These standard penalties will be reused throughout the paper, sometimes up to additive constants, which do not affect the minimizing assignments.

\section{LinkedIn's Queens and N-Queens problem}
\subsection{Problem definition}
\begin{figure}
    \centering
    \includegraphics[width=\linewidth]{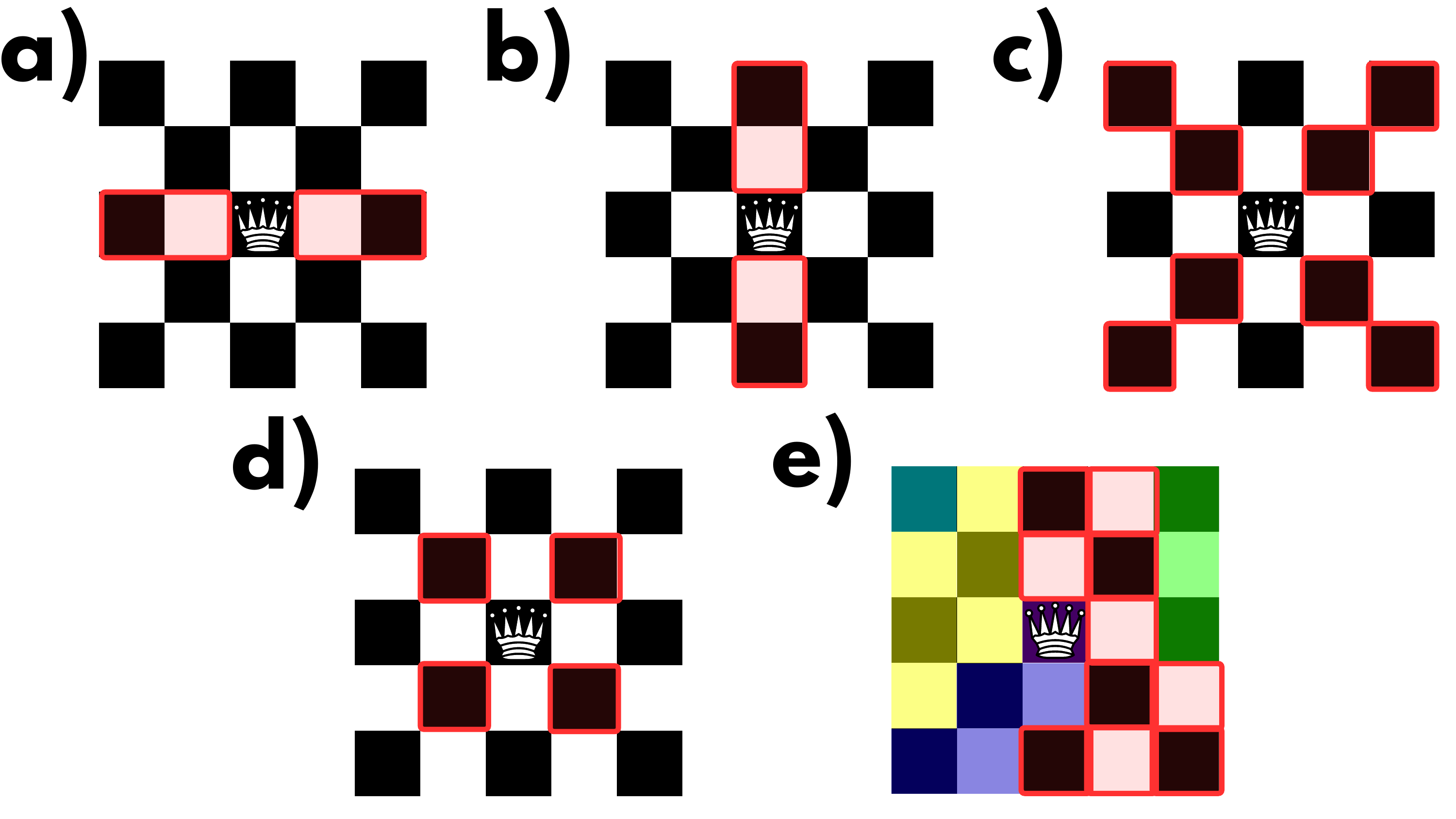}
    \caption{Constraints on a $5\times 5$ problem, such that no queens can be placed in any of the cells marked in red. a) Column constraint, b) Row constraint, c) Diagonal constraint on N-queens, d) Diagonal constraint on LQueens, e) Region constraint.}
    \label{fig: Restrictions}
\end{figure}

The original problem to be solved in the N-queens consists of placing $N$ queens on a chessboard of dimensions $N\times N$ in such a way that none of them threatens the others. That is, no queen can be in the same row or column as another queen, nor be diagonal to another queen. The constraints are shown in Fig. \ref{fig: Restrictions}.

To formalize it mathematically, we say that the state of the board is determined by a binary matrix $x$ whose elements $x_{ij}$ determine the state of the cell in row $i$ and column $j$, so that if it is 0 there is no queen in that cell, and if it is 1 there is. Therefore, the restrictions are 

\begin{itemize}
    \item Rows condition: there can only be one queen per row (Fig. \ref{fig: Restrictions} a):$$\sum_{j=1}^N x_{ij}=1\quad \forall i\in[1,N].$$
    \item Columns condition: there can only be one queen per column (Fig. \ref{fig: Restrictions} b): $$\sum_{i=1}^N x_{ij}=1\quad \forall j\in[1,N].$$
    \item Diagonal condition: there cannot be two queens on the same attacking diagonal (Fig. \ref{fig: Restrictions} c), namely
    $$x_{ij}+x_{kl}\leq 1\quad \forall (i,j)\neq (k,l)\ \text{with}\ |i-k|=|j-l|.$$
    We will use $D_{ij}$ only to denote the cells diagonally aligned with $(i,j)$, as an indexing device for these pairwise conflicts, not as a single region on which one imposes an at-most-one condition over the union of both diagonals through $(i,j)$.
\end{itemize}

The condition that there be $N$ queens is implicit in the first two restrictions.

LinkedIn's version of the game is slightly different. While it retains the first two restrictions, the third restriction becomes softer, making it so that queens can now not be diagonal only in the case where they are adjacent, which translates to
\begin{itemize}
    \item Diagonal condition: there cannot be more than one queen in the same adjacent diagonal (Fig. \ref{fig: Restrictions} d): $x_{ij}+X_{i+1,j+1}+X_{i-1,j+1}+X_{i+1,j-1}+X_{i-1,j-1} \leq 1\quad \forall i,j\in[1,N]$, being $X_{kl}=x_{kl}$ if $k,l\in[1,N]$ and $0$  otherwise.
\end{itemize}
Another modification in the LinkedIn version is the region restriction, which imposes on us that we have to place one and only one queen for each one of the $N_R=N$ colored regions, which translates to
\begin{itemize}
    \item Region condition: there can only be one queen per region (Fig. \ref{fig: Restrictions} e): $\sum_{(i,j)\in R_k} x_{ij}=1\quad \forall k\in[1,N_R]$, being $R_k$ the cells belonging to the $k$-th region.
\end{itemize}

An additional condition in this case is that there are a number $N_I$ of queens already placed in cells $A_I$, as an initial condition. This can be addressed by forcing the variables associated with these cells to be forcibly 1, which leads to all other conflicting variables in the same row, in the same column, in the same region and in the adjacent diagonals to be forcibly 0. This translates to
\begin{itemize}
    \item Initial condition: there are $N_I$ pre-placed queens:
    \begin{itemize}
        \item Cell initialized: $x_{lm}=1\quad \forall (l,m)\in A_I$
        \item Row: $x_{lj}=0\quad \forall j\in[1,N]$ if $j\neq m$ and  $\forall (l,m)\in A_I$
        \item Column: $x_{im}=0\quad \forall i\in[1,N]$ if $i\neq l$ and  $\forall (l,m)\in A_I$
        \item Region: $x_{ij}=0\quad \forall (i,j)\in R_k $ if $(i,j)\neq (l,m )$ and $ (l,m)\in R_k$ $\forall (l,m)\in A_I$
        \item Adjacent diagonals: $x_{ij}=0$ for every cell $(i,j)$ with $|i-l|=|j-m|=1$ and $\forall (l,m)\in A_I$.
    \end{itemize}
\end{itemize}

\begin{figure}
    \centering
    \includegraphics[width=\linewidth]{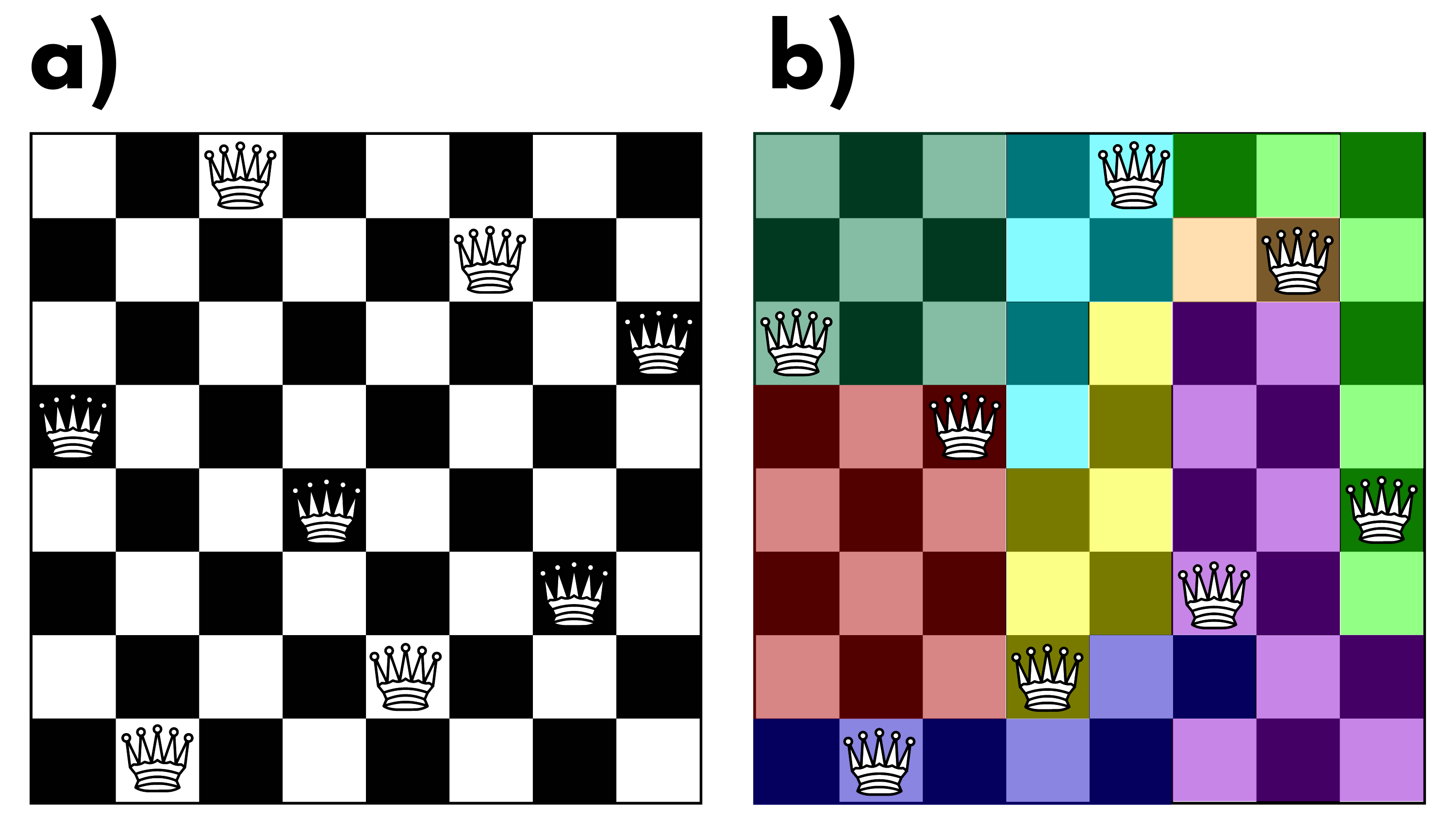}
    \caption{a) Solution of an 8-queen problem, b) Solution of an 8-queen problem from LinkedIn}
    \label{fig: General Table}
\end{figure}

In Fig. \ref{fig: General Table} a) we can see an example solution in the original N-queens, while in Fig. \ref{fig: General Table} b) we can see an example solution of the LinkedIn Queens.

\subsection{QUBO resolution}
For the QUBO formulation of the problem, we need to be able to put the constraints as a quadratic form, making the constraints work for the variables in pairs. We first tackle the standard N-queens and then the LinkedIn queens.

For the N-queens we need all the $N^2$ variables, so we need $N^2$ logical qubits. The first constraint is an equality, which we can impose with a quadratic term of the form
\begin{equation}\label{eq: rows cond}
    Q_r=\sum_{i=1}^N \left(1 -\sum_{j=1}^N x_{ij}\right)^2.
\end{equation}
This term has its minimum, equal to zero, when for each row there is only one variable equal to 1, so that all solutions that do not meet this restriction have a higher cost.

For the column condition, the term is analogous, interchanging the summands in $i$ and in $j$
\begin{equation}\label{eq: columns cond}
    Q_c=\sum_{j=1}^N \left(1 -\sum_{i=1}^N x_{ij}\right)^2.
\end{equation}

For the restriction of the diagonals we use a conflict-graph term of the type
\begin{equation}\label{eq: diagonals cond}
    Q_d=\sum_{\{u,v\}\in E_{\mathrm{diag}}} x_u x_v
    =\sum_{i,j=1}^{N-1,N}\sum_{(k,l)\in D'_{i,j}} x_{ij}x_{kl},
\end{equation}
where $E_{\mathrm{diag}}$ is the set of unordered pairs of cells lying on a common attacking diagonal. If we keep the notation $D'_{i,j}$, it must be understood as the set of cells below $(i,j)$ that are in diagonal conflict with it, so that the sum counts conflicting pairs exactly once. In particular, we are not imposing an at-most-one condition on an arbitrary union of the two diagonals through a cell, which would be stronger than the actual N-queens rule.

The total QUBO formulation is
\begin{equation}
    Q = Q_r + Q_c + Q_d =\sum_{i=1}^N \left(1 -\sum_{j=1}^N x_{ij}\right)^2 + \sum_{j=1}^N \left(1 -\sum_{i=1}^N x_{ij}\right)^2 +\sum_{i,j=1}^{N-1,N}\sum_{(k,l)\in D'_{i,j}} x_{ij}x_{kl}.
\end{equation}

All valid solutions to this problem have a cost $Q=0$, while all solutions that do not satisfy the constraints have a cost $Q>0$.

For the LQueens the formulation is similar. The first adjustment we have to make is to consider what are the fixed variables due to the initial condition. These are removed directly from the optimisation problem, but we keep the notation of the celles for clarity. That is, if we have the $(3,3)$ cell, the variable referring to the $(4,3)$ cell will still be $x_{4,3}$, since we know that there really is no $x_{3,3}$. So we will have a number of variables equal to $N^2$ if there are no queens at the beginning, and less if we add queens in the initial condition.

Let's define for simplicity the sets of cells:
\begin{itemize}
    \item $A_C$: contains the cells that have not been fixed by the initial condition (all variables to be optimised).
    \item $A'_C$: contains the cells of $A_C$ that are not in the last row of $A_{r}$.
    \item $A_R$: contains the indexes of all but one of the regions that have at least one cell in $A_C$.
\end{itemize}

and the sets of indexes:
\begin{itemize}
    \item $A_{r}$: contains all rows $i$ containing at least one cell in $A_C$.
    \item $A_{c}$: contains all columns $j$ containing at least one cell in $A_C$.
    \item $A_{c}(i)$: the set of $j$ such that $(i,j)\in A_C$.
    \item $A_{r}(j)$: the set of $i$ such that $(i,j)\in A_C$.
\end{itemize}

Thus, the term of the rows is
\begin{equation}
    Q_r=\sum_{i\in A_{r}} \left(1 -\sum_{j\in A_{c}(i)} x_{ij}\right)^2.
\end{equation}

The term of the columns is
\begin{equation}
    Q_c=\sum_{j\in A_{c}} \left(1 -\sum_{i\in A_{r}(j)} x_{ij}\right)^2.
\end{equation}
In both cases, the only modification with respect to the N-queens has been to eliminate from the summations the rows and columns without cells to be determined and the cells already fixed.

For the diagonal condition in this case, we again use a pairwise conflict term, now restricted to adjacent diagonals. Thus $D'_{ij}$ contains only the diagonally adjacent cells in rows below $i$ that remain unfixed after preprocessing. Because of the initial condition, we have to construct the term as

\begin{equation}
    Q_d=\sum_{(i,j)\in A'_C}\sum_{(k,l)\in D'_{i,j}} x_{ij}x_{kl},
\end{equation}
so that every remaining adjacent-diagonal conflict is still counted exactly once. In this case we only consider cells that are not fixed.

Now, for the region condition, we add a term
\begin{equation}\label{eq: region cond}
    Q_g = \sum_{k\in A_R} \left(1-\sum_{(i,j)\in R_k|(i,j)\in A_C} x_{ij}\right)^2.
\end{equation}
ensuring that the minimum value 0 in the case that each region has only one queen in each region. We consider only $N-N_I-1$ regions, since if there is a queen in each of these, the remaining queen must necessarily be in the remaining region, and it can only be one, because of the above restrictions.

Therefore, the total QUBO formulation in this case is
\begin{align}
    Q =& Q_r + Q_c + Q_d + Q_g =\sum_{i\in A_{r}} \left(1 -\sum_{j\in A_{c}(i)} x_{ij}\right)^2 +\sum_{j\in A_{c}} \left(1 -\sum_{i\in A_{r}(j)} x_{ij}\right)^2 +\nonumber\\
    +&\sum_{(i,j)\in A'_C}\sum_{(k,l)\in D'_{i,j}} x_{ij}x_{kl}+\sum_{k\in A_R} \left(1-\sum_{(i,j)\in R_k|(i,j)\in A_C} x_{ij}\right)^2.
\end{align}

As before, all valid solutions to this problem have a cost $Q=0$, while all solutions that do not satisfy the constraints have a cost $Q>0$.

\begin{proposition}
After preprocessing the pre-placed queens by fixing $x_{lm}=1$ for $(l,m)\in A_I$ and fixing to $0$ every remaining cell in the same row, column, region or adjacent diagonal, the minimum-energy states of the reduced LQueens QUBO coincide with the solutions of the original instance restricted to the non-fixed variables.
\end{proposition}

\begin{proof}
Every constraint involving a pre-placed queen is already satisfied by the preprocessing, because every cell that could violate such a constraint has been fixed to $0$. Hence all removed terms are constant on the reduced instance. Conversely, any assignment of the remaining variables violates the original puzzle if and only if it violates one of the surviving row, column, adjacent-diagonal or region constraints, which are precisely the terms kept in the reduced QUBO.
\end{proof}

\subsection{General cases}
In the previous section we have solved the particular cases of N-queens and the LQueens. However, we can generalise this formulation to solve more complex problems. As we have seen, each constraint has an associated QUBO term, and the initial condition determines the number of variables we have in the problem. Therefore, let's determine the terms of the QUBO problem for each constraint. Since the constraints may be incompatible with each other, this formulation may also be used to test whether a certain generalized problem has a solution that satisfies its constraints.

\begin{figure}
    \centering
    \includegraphics[width=\linewidth]{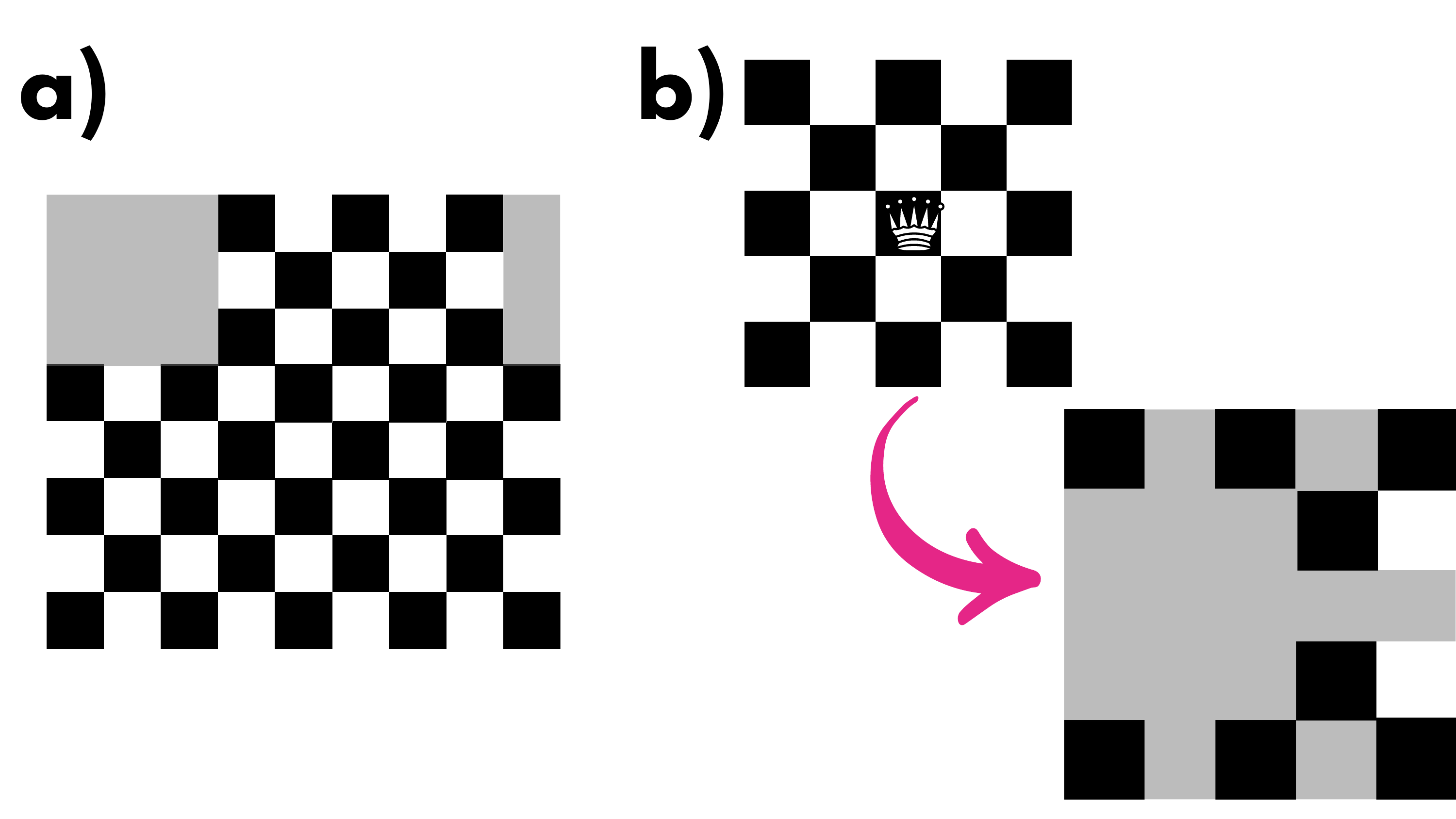}
    \caption{a) Irregular shaped board. b) Regular board with initial condition and its equivalent irregular board without initial condition. The cells in grey are squares that are not on the board.}
    \label{fig: Irregular}
\end{figure}

As the initial condition only fixes the set of variables of the problem, in order to make the notation clear and without loss of generality, we will not add it. We will consider a rectangular board of dimensions $N\times M$. If we wanted irregular shapes, we would only have to create a rectangular board such that it can contain all the cells of the wanted one, and fix all cells outside the desired board to zero. In the same way, we can reduce a problem with an initial condition to a problem without an initial condition with a board with only the cells that are not fixed by the initialization, and adjusting the number of queens in each region according to the number of queens already existing in that region by the initial condition. We can see an example of an irregular board in Fig. \ref{fig: Irregular} a), and how to translate a problem with initial condition to one without initial condition by changing the shape of the board in Fig. \ref{fig: Irregular} b).

\subsubsection{Number of queens per row/column}
The first generalised constraint to consider is the following. Given a board of dimension $N\times M$, there can only be $r_i$ queens in the $i$-th row. As can be understood, the row restriction in the cases studied above is the particular case where $r_i=1$ for all rows.

The term that implements this constraint is a generalisation of \eqref{eq: rows cond}
\begin{equation}\label{eq: general rows}
    Q_r=\sum_{i=1}^N \left(r_i -\sum_{j=1}^M x_{ij}\right)^2.
\end{equation}

The next constraint to implement is, given a board of dimension $N\times M$, there can only be $c_j$ queens in the $j$-th column. Once again, the column restriction in the cases studied above is the particular case where $c_j=1$ for all columns.

The term that implements this constraint is a generalisation of \eqref{eq: columns cond}
\begin{equation}\label{eq: general columns}
    Q_c=\sum_{j=1}^M \left(c_j -\sum_{i=1}^N x_{ij}\right)^2.
\end{equation}

In the case of an initial condition, these terms have to be modified both by eliminating the cells in which there can no longer be any queens (because their column and row already contain the required number of queens) and by reducing the number of queens required in the row and column in which we have already placed a queen. That is, if in row 3 we have placed a queen by initial condition, the number of queens available for the rest of cells in the row will be $r_3-1$, so in the QUBO we will have to reduce each $r_i$ and $c_j$ with the number of queens we initially placed in row $i$ and column $j$, respectively.

\subsubsection{Distant diagonals constraint}
In the case of N-queens, we required that if there was a queen on a cell, there could not be any more queens on any of its diagonals until the end of the board, whereas in LQueens it is only required for cells on the adjacent diagonal. To generalise this case, we say that we have a diagonal condition of distance $d_{ij}$ if in case there is a queen on the cell $(i,j)$ there cannot be any queen on any cell located on any of its diagonals at a distance equal to or less than $d_{ij}$ cells. In this way, not only do we allow diagonals of different lengths to be considered, but also that this depends on the cell.

The case of the N-queens is the particular case of $d_{ij}=\max(N,M)\quad \forall i\in [1,N],j\in[1,M]$, and LQueens that of $d_{ij}=1\quad \forall i\in [1,N],j\in[1,M]$.

The term that implements this remains a conflict-graph term. Let $E_d$ be the set of unordered pairs of cells such that one of them lies on one of the allowed diagonals of the other and at the required distance. Then the penalty is obtained by summing one product per conflicting pair. If one prefers to keep a set notation, $D''_{ij}$ should only enumerate the cells that conflict with $(i,j)$, not define a single region with an at-most-one condition over a union of diagonals.

\begin{equation}\label{eq: general diagonals}
    Q_d=\sum_{\{u,v\}\in E_d} x_u x_v.
\end{equation}

Given this abstract formulation, this condition can be generalized so that the 4 diagonals do not have the same distance, that instead of being continuous lines they can be skipped lines, or that instead of being 4 diagonals they are $n_d$ attack lines with different slopes. In every case, what matters is the underlying conflict graph, not an arbitrary region obtained by merging unrelated lines.

\subsubsection{Number of queens per region}
In the case of LQueens we imposed that, for a set of $N$ non-overlapping regions, in each of them there should be one and only one queen. However, we can generalize it so that regions can overlap, since it will only affect the definition of the $R_k$, and that instead of there having to be 1 queen per region, there must be $q_k$ queens in the $k$-th region. 

This term will be a generalization of \eqref{eq: region cond} for a number $N_R$ of regions.
\begin{equation}
    Q_g = \sum_{k=1}^{N_R} \left(q_k-\sum_{(i,j)\in R_k} x_{ij}\right)^2.
\end{equation}

As can be seen, this term is a generalization of both the row term and the column term, since we can make one region per row and one region per column, and it would result in the same terms.

If we had an initial condition, this term would have to be corrected by subtracting from $q_k$ the number of queens we have placed at the beginning in the $k$-th region, eliminating the cells that have already been fixed or that are in regions in which there can be no more queens.

Therefore, a generalized queens problem with a board of dimension $N\times M$ with no initial condition can be expressed by the two terms

\begin{equation}\label{eq: general 1}
    Q = Q_g + Q_d = \sum_{k=1}^{N_R+N+M} \left(q_k-\sum_{(i,j)\in R'_k} x_{ij}\right)^2+ \sum_{\{u,v\}\in E_d} x_u x_v,
\end{equation}
where the set of regions $R'$ contains the $N_R$ regions $R_k$, the $N$ rows and the $M$ columns, while $E_d$ is the conflict graph describing the diagonal-style attacks.

\subsubsection{Two possible numbers of queens per region}
A final generalization we can make is to create a constraint that is: given a board of dimension $N\times M$, in which there are $N_V$ regions that can be overlapping, there must be a number $v_k$ or $v_k+1$ of queens in region $k$ for each region. This constraint can be imposed by using a fractional term \cite{Fractionary} that places the minimum cost in the middle of both values
\begin{equation}\label{eq: two number region}
    Q_v = \sum_{k=1}^{N_V} \left(v_k+\frac{1}{2}-\sum_{(i,j)\in R_k} x_{ij}\right)^2.
\end{equation}
If we had an initial condition, we would have to apply the same correction as in the previous case.

The term \eqref{eq: two number region} is useful for genuine region constraints with two consecutive admissible occupancies. Diagonal attack constraints are instead better expressed through pairwise conflict graphs as in \eqref{eq: diagonals cond} and \eqref{eq: general diagonals}. We can rewrite the eq \eqref{eq: general 1}, with $N_{R}$ regions $R_k$ with a single possible number of queens and $N_{V}$ regions $V_k$ with two consecutive possible numbers of queens, as
\begin{equation}
    Q = Q_g + Q_v = \sum_{k=1}^{N_R} \left(q_k-\sum_{(i,j)\in R_k} x_{ij}\right)^2 +\sum_{k=1}^{N_V} \left(v_k+\frac{1}{2}-\sum_{(i,j)\in V_k} x_{ij}\right)^2.
\end{equation}

Finally, if we want to write the whole cost function as a single term, for $N_R$ regions $R_k$, we can define a value $t_k$, which will be 0 if the region $k$ can only have a single number of queens $q_k$ and 1 if the region $k$ can have two numbers of queens $q_k$ or $q_k+1$, such that
\begin{equation}
    Q = \sum_{k=1}^{N_T} \left(q_k+\frac{t_k}{2}-\sum_{(i,j)\in R_k} x_{ij}\right)^2.
\end{equation}
This compact expression summarizes only the region part of the objective. Whenever diagonal or line-of-sight attacks are present, the corresponding conflict-graph penalty must still be added separately. Moreover, for the terms with $t_k=1$, feasible assignments share a constant offset of $1/4$ per such region unless one uses the shifted form from \eqref{eq: prelim q qplus zero}.

\subsubsection{Toroidal board}
An interesting generalization is to consider a toroidal board, so that the pieces that pass through the upper part of the board exit through the lower part, those that pass through the right part exit through the left part and vice versa. To consider this problem, we must first impose that the side of the board has a minimum prime factor of 5, otherwise it will have no solution. We can also approach the case of a cylindrical board, in which this cyclic behavior will only occur when crossing the left and right sides.

\begin{figure}
    \centering
    \includegraphics[width=\linewidth]{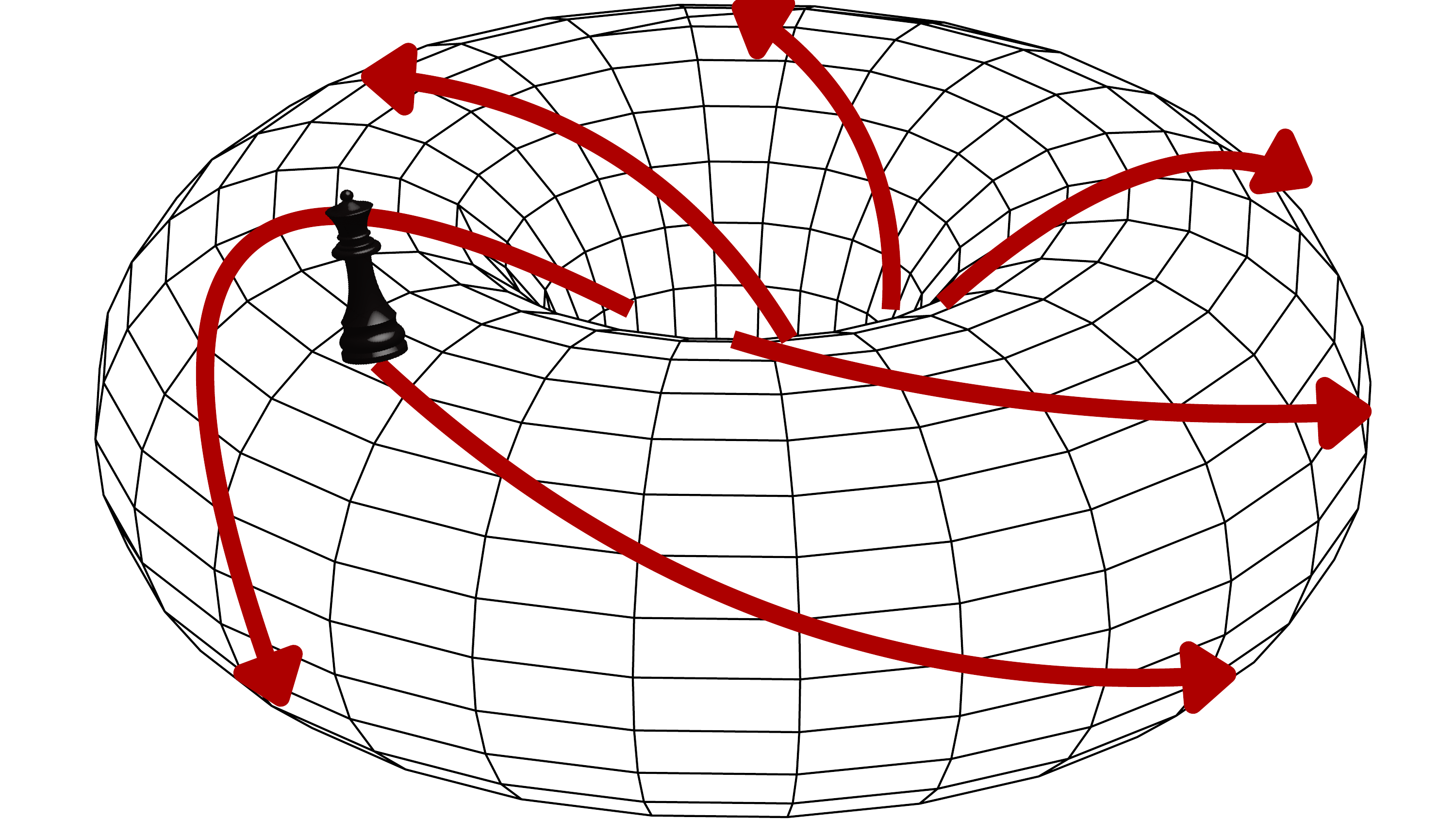}
    \caption{Toroidal board with a queen, where we mark in red what would be the squares that this queen could attack from only one of its 4 diagonals.}
    \label{fig: Toroid}
\end{figure}

To take into account this generalization, the terms of rows and columns will be exactly the same, as well as the term of the regions. The only term that will change will be that of the diagonals, since this, according to the length and shapes of the diagonals, will have to consider the cells reached after curving the boundary as many times as appropriate. Therefore, we will only have to change the definition of the $D'_{ij}$ to include all the necessary cells.

\subsection{Definition of generalised N-queens}
With the terms and generalisations we have created for the N-queens, we can define a fully general N-queens, and its associated QUBO, and particularise it to a number of interesting families of N-queens problems.

The general N-queens problem is defined as:

Given a board of arbitrary shape, of maximum height $N$ and maximum width $M$, on which there are placed $N_I$ initial queens in positions given by the set $A_I$, and divided into $N_R$ regions, which can overlap each other, such that in the $k$-th region there must be $q_k$ queens if $t_k=0$ and $q_k$ or $q_k+1$ queens if $t_k=1$, and we have initially placed $p_k$ queens, the objective will be to place all the possible queens on the board such that the given region restrictions and pairwise attack restrictions are satisfied.

The QUBO formulation of this problem is given in general form as
\begin{equation}
    Q = \sum_{k=1}^{N_R} \left(q_k-p_k+\frac{t_k}{2}-\sum_{(i,j)\in R_k} x_{ij}\right)^2 + \sum_{\{u,v\}\in E_d} x_u x_v,
\end{equation}
where $R_k$ is the set of cells included in the $k$-th region that have not been set in the initial condition and that do not belong to a region where the number of queens available after the initial queens have been placed is 0, and $E_d$ is the chosen conflict graph for the diagonal or line-of-sight restrictions. If $t_k=1$, the corresponding feasible region contributes a constant offset of $1/4$ unless one uses the shifted form from \eqref{eq: prelim q qplus zero}.

We will propose the form of two particular versions of this problem.

\subsubsection{N-queens with regions}
In this case the problem is a combination of the original N-queens with the LQueens, adding to the N-queens the $N_R$ regions in which there can only be one queen.

In this case, the QUBO formulation is
\begin{align}
    Q =& Q_r + Q_c + Q_d + Q_g =\sum_{i=1}^N \left(1 -\sum_{j=1}^N x_{ij}\right)^2 + \sum_{j=1}^N \left(1 -\sum_{i=1}^N x_{ij}\right)^2 +\nonumber\\
    +&\sum_{i,j}^{N-1,N}\sum_{(k,l)\in D'_{i,j}} x_{ij}x_{kl}+\sum_{k=1}^{N_R} \left(1-\sum_{(i,j)\in R_k} x_{ij}\right)^2.
\end{align}

\subsubsection{N-queens with soft regions}
This is the case of N-queens with regions, but allowing each region to have a queen or not.

In this case, the QUBO formulation is
\begin{align}
    Q =& Q_r + Q_c + Q_d + Q_v =\sum_{i=1}^N \left(1 -\sum_{j=1}^N x_{ij}\right)^2 + \sum_{j=1}^N \left(1 -\sum_{i=1}^N x_{ij}\right)^2 +\nonumber\\
    +&\sum_{i,j}^{N-1,N}\sum_{(k,l)\in D'_{i,j}} x_{ij}x_{kl}+\sum_{k=1}^{N_R} \left(\frac{1}{2}-\sum_{(i,j)\in R_k} x_{ij}\right)^2.
\end{align}
Here the feasible soft-region assignments share a constant offset of $N_R/4$ because of the fractional terms.

Another equivalent formulation would be to change the last term to one similar to that of the diagonals

\begin{align}
    Q =&\sum_{i=1}^N \left(1 -\sum_{j=1}^N x_{ij}\right)^2 + \sum_{j=1}^N \left(1 -\sum_{i=1}^N x_{ij}\right)^2 +\nonumber\\
    +&\sum_{i,j}^{N-1,N}\sum_{(k,l)\in D'_{i,j}} x_{ij}x_{kl}+\sum_{k=1}^{N_R}\sum_{(i,j)\in R_k}\sum_{(l,m)\in R_k|(l,m)\neq (i,j)} x_{ij}x_{lm}.
\end{align}

\subsection{Tents \& Trees}
An interesting game related to these generalizations is Tents \& Trees. This game is formulated as follows:
Given a board of dimensions $N\times N$ cells, with $N_t$ trees placed at positions $A_t$, the objective is to place a series of tents such that the following constraints are satisfied:
\begin{itemize}
    \item Each tree will have an associated tent, in a cell vertically or horizontally adjacent to it.
    \item No two tents can be adjacent to each other, not even diagonally.
    \item Each row $i$ and column $j$ has to have a number $N_{r,i}$ and $N_{c,j}$ of associated tents.
\end{itemize}

If one directly adapts the regional formalism, each tree neighborhood can be treated as a region that should contain one or two tents, and the no-touch condition can be enforced through pairwise conflicts between king-adjacent cells. This gives a useful simplified relaxation, but it is not exact in general because overlapping neighborhoods do not enforce a one-to-one assignment between trees and tents.

For an exact formulation, let $F$ be the set of free cells, excluding the tree cells themselves. For each $c\in F$, introduce a binary variable $x_c$ indicating whether there is a tent at cell $c$. For each tree $t$ and each free cell $c$ orthogonally adjacent to $t$, introduce a binary assignment variable $y_{t,c}$, and denote by $N(t)\subseteq F$ the set of free cells orthogonally adjacent to $t$. Let $E_{\mathrm{king}}$ be the conflict graph on $F$ connecting two cells when they are adjacent orthogonally or diagonally.

The exact penalties are then
\begin{equation}
    Q_{\mathrm{tree}}
    =
    A\sum_{t}\left(\sum_{c\in N(t)} y_{t,c} - 1\right)^2,
\end{equation}
\begin{equation}
    Q_{\mathrm{assign}}
    =
    A\sum_{c\in F}\left(x_c-\sum_{t:\, c\in N(t)} y_{t,c}\right)^2,
\end{equation}
\begin{equation}
    Q_{\mathrm{touch}}
    =
    A\sum_{\{c,c'\}\in E_{\mathrm{king}}} x_c x_{c'},
\end{equation}
\begin{equation}
    Q_{\mathrm{row}}
    =
    A\sum_{i=1}^{N}\left(\sum_{j:\, (i,j)\in F} x_{ij} - N_{r,i}\right)^2,
\end{equation}
\begin{equation}
    Q_{\mathrm{col}}
    =
    A\sum_{j=1}^{N}\left(\sum_{i:\, (i,j)\in F} x_{ij} - N_{c,j}\right)^2.
\end{equation}
If a particular puzzle does not specify row or column counts, the corresponding term can simply be omitted.

Therefore, an exact QUBO for Tents \& Trees is
\begin{equation}
    Q_{Tents}^{\mathrm{exact}}
    =
    Q_{\mathrm{tree}}
    + Q_{\mathrm{assign}}
    + Q_{\mathrm{touch}}
    + Q_{\mathrm{row}}
    + Q_{\mathrm{col}}.
\end{equation}

\begin{proposition}
The zero-energy states of $Q_{Tents}^{\mathrm{exact}}$ are exactly the standard solutions of Tents \& Trees.
\end{proposition}

\begin{proof}
$Q_{\mathrm{tree}}$ gives exactly one assigned tent to each tree. $Q_{\mathrm{assign}}$ guarantees that every occupied tent cell is assigned to exactly one adjacent tree and that no assignment is made to a cell without a tent. $Q_{\mathrm{touch}}$ forbids orthogonal and diagonal contact between tents, while $Q_{\mathrm{row}}$ and $Q_{\mathrm{col}}$ impose the row and column counts. Conversely, every valid puzzle solution induces a unique compatible assignment of each tent to its adjacent tree, so all five penalties vanish.
\end{proof}

\subsection{Chess Piece Problems}
As can be seen, this formulation allows, not only to solve the case for the movement of queens, but also for the movement of any type of piece.

Therefore, we can generalise the problem to the chess piece problem, which will consist of trying to place as many pieces as possible on a board of maximum dimensions $N\times M$ so that no piece threatens another, with the extra restriction that there can only be one piece per coloured region and each cell can only be empty or have a specific type of piece, which can vary between cells.

\begin{figure}
    \centering
    \includegraphics[width=\linewidth]{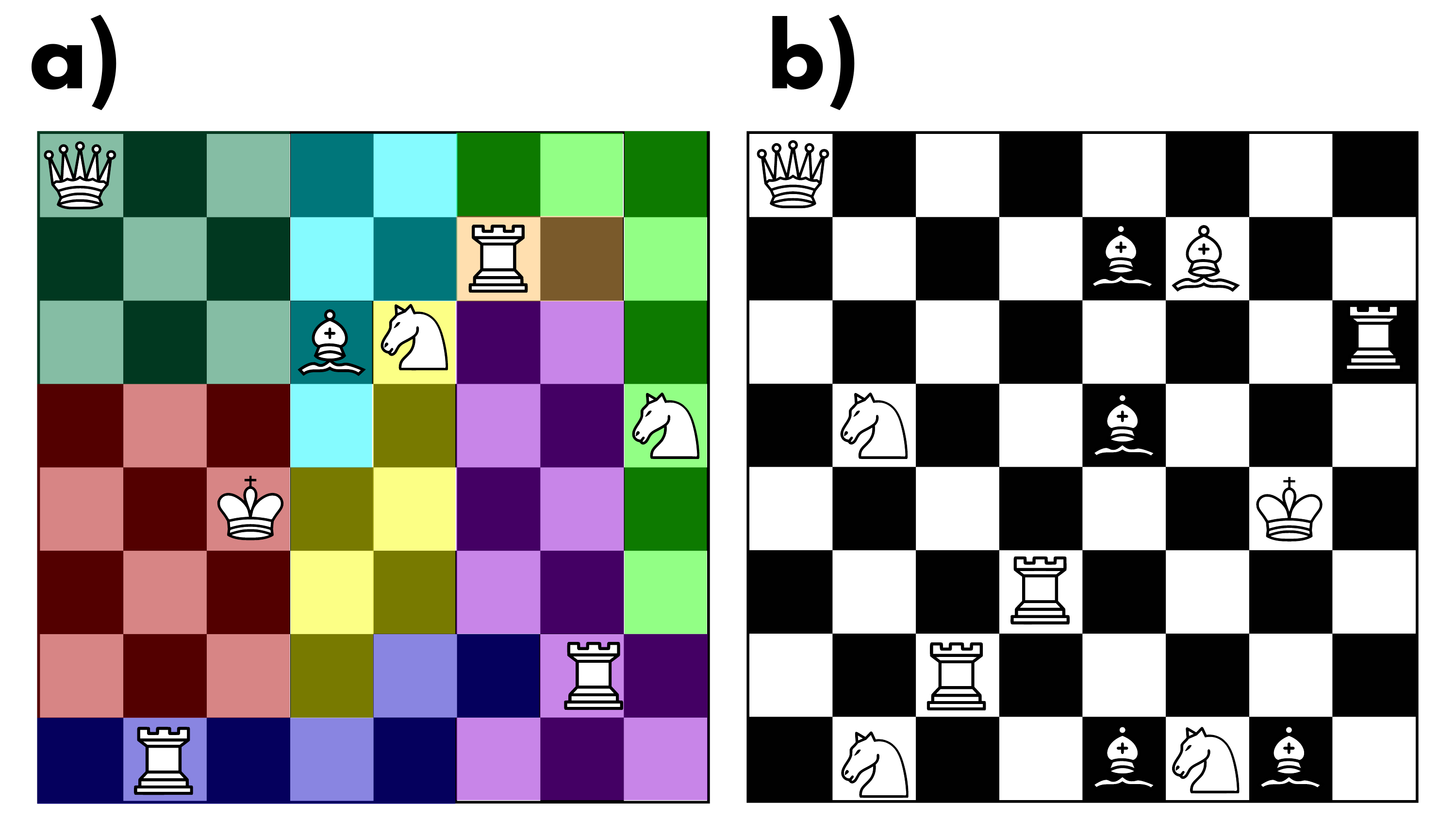}
    \caption{a) Solution of a rectangular Coloured Chess Piece Problem, b) Solution of a rectangular Max Chess Pieces Problem}
    \label{fig: Example_pieces}
\end{figure}

The Coloured Chess Piece Problem is:

Given a board of arbitrary shape, of maximum height $N$ and maximum width $M$, on which there are placed a number $N_p$ of initial pieces in positions given by the set $A_I$, and divided into $N_R$ regions, which cannot overlap each other, such that in each region there must be only one piece, the objective will be to place all the possible pieces on the board such that no piece threatens another. Each cell $(i,j)$ can be empty or have the piece of type $P_{ij}$, with possible moves $M_{P_{ij}}$, and can reach cells $C\left(M_{P_{ij}}\right)_{ij}$. A solution of this problem is represented in Fig. \ref{fig: Example_pieces} a.

As in the case of the queens, we will only have to eliminate the cells in which we have set the variables when initializing. That is,
in which we have placed the pieces ($(i,j)\in A_I$), in the regions in which we have placed them ($(l,m)\in R_k | (i,j)\in R_k, \forall (i,j) \in A_I$) and in the cells threatened by the placed pieces ($(l,m)\in C\left(M_{P_{ij}}\right)_{ij} \forall (i,j) \in A_I$). These cells can only be empty, so their variables will not be considered in the optimization, although we can place them in the QUBO formulation, since such values of 0 will not alter the summations. We will define the $R'_k$ regions as the $N'_R$ regions in which no pieces have been placed at the beginning.

This problem has the restrictions:
\begin{itemize}
    \item Coloured Regions: There can be only one piece per coloured region:
    \begin{equation*}
        \sum_{(i,j)\in R'_k} x_{ij} = 1\quad \forall k\in [1,N'_R]
    \end{equation*}
    \item No threat: No piece can be threatening another piece:
    \begin{equation*}
        \sum_{(l,m)\in C\left(M_{P_{ij}}\right)_{ij}} x_{lm}= 0 \quad \forall (i,j) | x_{ij} = 1
    \end{equation*}
\end{itemize}

To implement these restrictions, we only need a coloured term

\begin{equation}
    Q_g = \sum_{k=1}^{N'_R} \left( 1 - \sum_{(i,j)\in R'_k} x_{ij}\right)^2
\end{equation}

and a threat term
\begin{equation}
    Q_t = \sum_{i=1}^{N}\sum_{j=1}^{M} \sum_{(l,m)\in C\left(M_{P_{ij}}\right)_{ij}}x_{ij} x_{lm}.
\end{equation}

The total QUBO formulation is
\begin{equation}
    Q = Q_g + Q_t = \sum_{k=1}^{N'_R} \left( 1 - \sum_{(i,j)\in R'_k} x_{ij}\right)^2 + \sum_{i=1}^{N}\sum_{j=1}^{M} \sum_{(l,m)\in C\left(M_{P_{ij}}\right)_{ij}}x_{ij} x_{lm}.
\end{equation}
Thus, a board that meets the constraints will be one with zero energy.

The other problem we can solve is the Max Chess Pieces Problem, which consists of placing as many pieces as possible on the board so that none of them threatens another, and there can be only one type per cell. We will define it as:

Given a board of arbitrary shape, of maximum height $N$ and maximum width $M$, on which there are placed a number $N_p$ of initial pieces in positions given by the set $A_I$, the objective will be to place the maximum number possible of pieces on the board such that no piece threatens another. Each cell $(i,j)$ can be empty or have the piece of type $P_{ij}$, with possible moves $M_{P_{ij}}$, and can reach cells $C\left(M_{P_{ij}}\right)_{ij}$. A solution of this problem is represented in Fig. \ref{fig: Example_pieces} b.

Since we only keep the threat constraint, we eliminate the coloured term and add a reward term that favors occupied cells. In weighted form, let $w_{ij}\geq 0$ be the reward for placing a piece in cell $(i,j)$, and define
\begin{equation}
    Q_n = -\sum_{i=1}^{N}\sum_{j=1}^{M} w_{ij}x_{ij}.
\end{equation}
The unweighted case corresponds to $w_{ij}=1$ for all cells.

Therefore, the total QUBO is
\begin{equation}
    Q = Q_n + \lambda Q_t = -\sum_{i=1}^{N}\sum_{j=1}^{M} w_{ij}x_{ij} + \lambda \sum_{i=1}^{N}\sum_{j=1}^{M} \sum_{(l,m)\in C\left(M_{P_{ij}}\right)_{ij}}x_{ij} x_{lm}.
\end{equation}

\begin{proposition}
If $\lambda>\max_{i,j} w_{ij}$, then no global minimizer of the Max Chess Pieces QUBO contains a conflicting pair of pieces. In particular, in the unweighted case it is sufficient to take $\lambda>1$.
\end{proposition}

\begin{proof}
Assume a minimizer contains a conflicting pair. Then there exists an occupied cell $z=(i,j)$ belonging to at least one conflict edge. If we switch $x_z$ from $1$ to $0$, the reward term increases by at most $w_{ij}\leq \max_{i,j}w_{ij}$, while the conflict term decreases by at least $\lambda$ because at least one conflicting product disappears. Since $\lambda>\max_{i,j}w_{ij}$, the total energy decreases, contradicting optimality.
\end{proof}

\section{Takuzu and LinkedIn's Tango Problems}
\subsection{Problem definition}
The problem to be solved in Takuzu is the following:

Given a square board of dimension $N\times M$ cells, where each cell can contain only one 0 or 1, initially with $N_{I0}$ zeros placed at positions $A_{I0}$ and $N_{I1}$ ones placed at positions $A_{I1}$, we have to determine for each cell whether there should be a zero or a one. The constraints are as follows:
\begin{itemize}
    \item In each row and column there must be as many zeros as ones.
    \item There cannot be more than two zeros or ones in a row vertically or horizontally.
    \item No two rows and no two columns can have the same combination of zeros and ones.
\end{itemize}

\begin{figure}
    \centering
    \includegraphics[width=\linewidth]{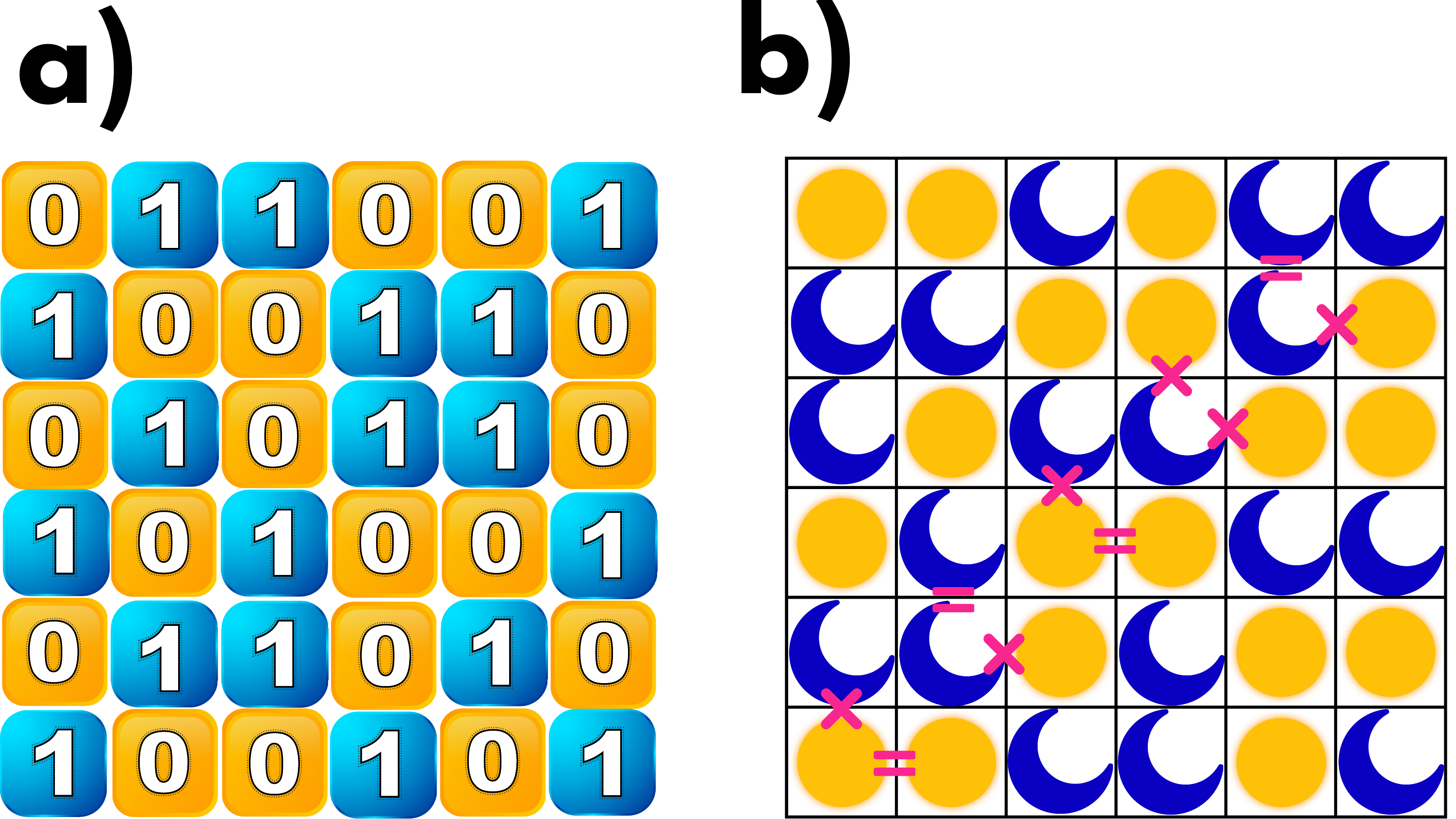}
    \caption{Solution for a problem a) Takuzu/0h h1 $6\times 6$, b) Tango}
    \label{fig: Solutions}
\end{figure}

A solution to this problem can be seen in Fig. \ref{fig: Solutions} a.

The Tango problem is very similar, but adding additional constraints and removing the restriction of repeating rows or columns. It is expressed as follows:

Given a square board of dimension $6\times 6$ cells, where each cell can contain only one sun icon or one moon icon, initially with $N_{IS}$ suns placed at positions $A_{IS}$ and $N_{IM}$ moons placed at positions $A_{IM}$, we have to determine for each cell whether there should be a sun or a moon. The constraints are as follows:
\begin{itemize}
    \item In each row and column there must be as many moons as suns.
    \item There cannot be more than two suns or moons in a row vertically or horizontally.
    \item There are cells between which there is an '=' symbol, which indicates that in both cells there must be the same icon.
    \item There are cells between which there is an 'x' symbol, which indicates that there must be a different icon in both cells.
\end{itemize}

A solution to this problem can be seen in Fig. \ref{fig: Solutions} b.

In this case, if we replace the suns by zeros and the moons by ones, and allow the board to be $N\times M$, we obtain a common local framework for both Takuzu and Tango. We call this local framework the Takuzu/Tango Problem (TTP): it contains the clues, the row and column balance conditions, the prohibition of three equal consecutive values, and optionally the '=' and 'x' relations. Classical Takuzu adds one more global condition, namely that no two rows and no two columns can coincide, whereas Tango does not require this global uniqueness.

\begin{figure}
    \centering
    \includegraphics[width=\linewidth]{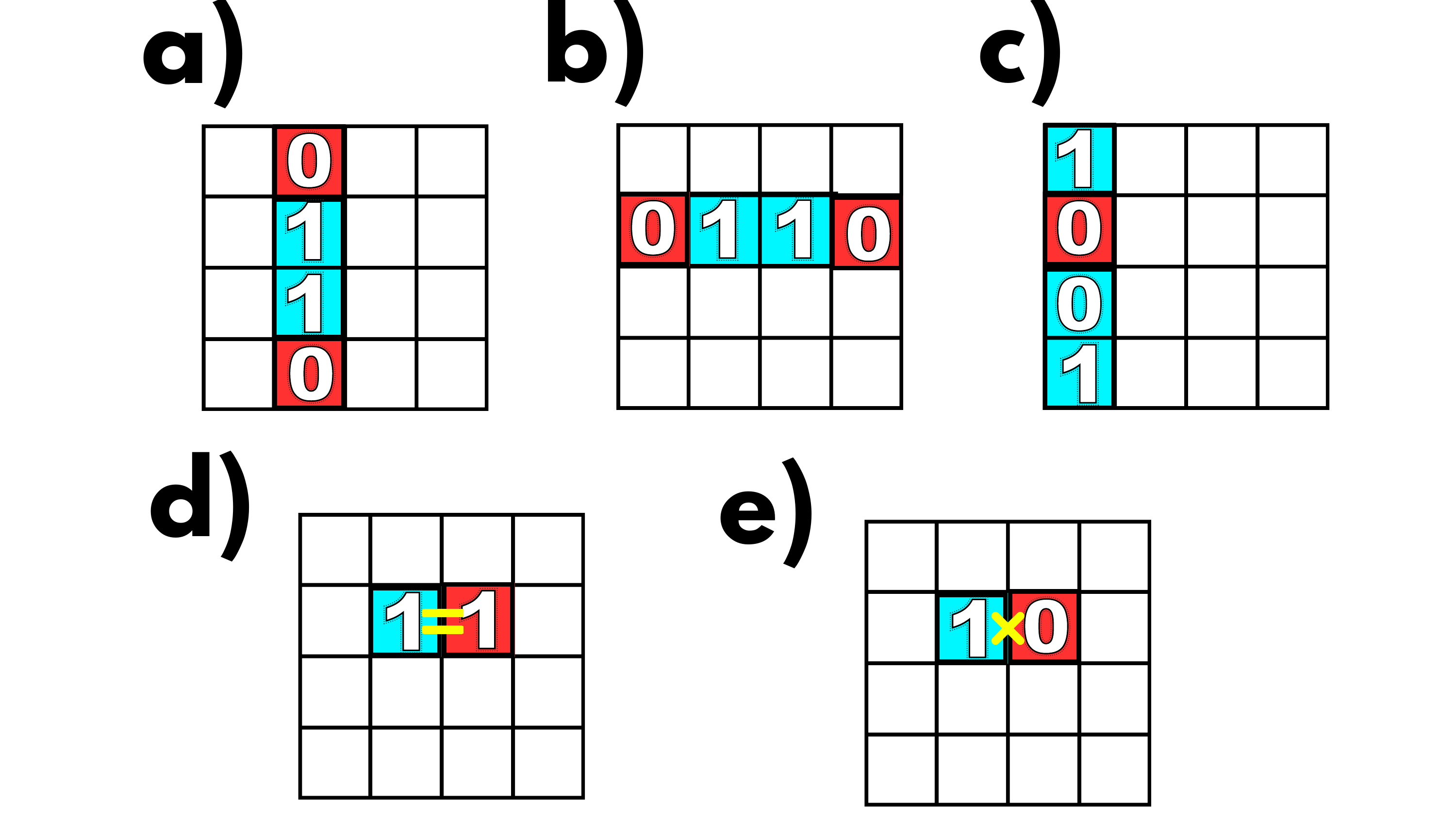}
    \caption{Boards with restrictions. The blue cells have the values previously set, while the red cells have their values determined by the constraints. a) Vertical repetition, b) Horizontal repetition, c) Regularity, d) Symbol '=', e) Symbol 'x'.}
    \label{fig: Restrictions 2}
\end{figure}

To formulate the local TTP, the state of a board is given by variables $x_{ij}\in\{0,1\}$, which indicate the value in the cell of row $i$ and column $j$. Therefore, the local constraints can be expressed as:
\begin{itemize}
    \item Row regularity: there must be as many zeros as ones in each row (Fig. \ref{fig: Restrictions 2} c):
    \begin{equation}
        \sum_{j=1}^{M} x_{ij} = \frac{M}{2}\quad \forall i\in [1,N].
    \end{equation}
    \item Column regularity: there must be as many zeros as ones in each column (Fig. \ref{fig: Restrictions 2} c):
    \begin{equation}
        \sum_{i=1}^{N} x_{ij} = \frac{N}{2}\quad \forall j\in [1,M].
    \end{equation}
    \item Vertical repetition condition: there cannot be more than two zeros or ones in a row vertically, so every 3-cell vertical window must contain one or two ones (Fig. \ref{fig: Restrictions 2} a):
    \begin{equation}
        \sum_{k=i}^{i+2} x_{kj} \in \{1,2\} \quad \forall i\in[1,N-2], j\in [1,M].
    \end{equation}
    \item Horizontal repetition condition: there cannot be more than two zeros or ones in a row horizontally, so every 3-cell horizontal window must contain one or two ones (Fig. \ref{fig: Restrictions 2} b):
    \begin{equation}
        \sum_{k=j}^{j+2} x_{ik}\in \{1,2\}\ \forall i\in[1,N],\ j\in [1,M-2].
    \end{equation}
    \item Equal sign condition: if there is an '=' symbol between two cells, both cells must have the same value.
    \item Cross sign condition: if there is an 'x' symbol between two cells, both cells must have different values.
\end{itemize}

For classical Takuzu, one must additionally impose the global non-repetition condition
\begin{equation}
    (x_{r1},\ldots,x_{rM})\neq (x_{s1},\ldots,x_{sM})\quad \forall 1\leq r<s\leq N,
\end{equation}
and analogously
\begin{equation}
    (x_{1c},\ldots,x_{Nc})\neq (x_{1t},\ldots,x_{Nt})\quad \forall 1\leq c<t\leq M.
\end{equation}
These global uniqueness constraints will be treated separately below: the local QUBO is exact for Tango and for the Takuzu/Tango variant without global uniqueness, while an exact QUBO for full Takuzu requires additional auxiliary variables.

\subsection{QUBO Formulation}
The local QUBO formulation is composed of terms that impose the clues, the balance conditions, the prohibition of three consecutive equal values and, when present, the parity signs. We first discuss how to reduce trivial variables. If we have zeros and ones already placed in positions $A_{I0}$ and $A_{I1}$, those cells do not need to be optimized. Moreover, if one of these fixed values is linked by an '=' or 'x' sign to another cell, then the adjacent cell can also be fixed immediately. The same happens when a row or column already contains as many zeros or ones as allowed by the balance condition, or when two equal fixed adjacent values force the third value in a length-3 window.

In unreduced form, the clue penalty can be written as
\begin{equation}
    Q_{\mathrm{clues}}= \sum_{(i,j)\in A_{I0}} x_{ij} + \sum_{(i,j)\in A_{I1}} (1-x_{ij}),
\end{equation}
which is a QUBO because the terms are linear in binary variables. After preprocessing, this term becomes a constant and may be omitted from the reduced optimization problem.

The horizontal repetition term is
\begin{equation}
    Q_{HR} = \sum_{i,j=1}^{N,M-2}\left(\frac{3}{2} - \sum_{k=j}^{j+2} x_{ik} \right)^2,
\end{equation}
and the vertical analogue is
\begin{equation}
    Q_{VR} = \sum_{i,j=1}^{N-2,M}\left(\frac{3}{2} - \sum_{k=i}^{i+2} x_{kj} \right)^2.
\end{equation}
These are fractional QUBO terms of the form discussed in \eqref{eq: prelim q qplus}: their minimizers are exactly the assignments for which each 3-cell window contains one or two ones, but each satisfied window contributes $1/4$ rather than $0$. If zero energy on each satisfied window is desired, one may subtract that constant or replace the term by the equivalent zero-based penalty from the preliminaries.

For the regularity conditions we use
\begin{equation}
    Q_{Hr} = \sum_{i=1}^{N}\left(\frac{M}{2} - \sum_{j=1}^{M} x_{ij} \right)^2,
\end{equation}
\begin{equation}
    Q_{Vr} = \sum_{j=1}^{M}\left(\frac{N}{2} - \sum_{i=1}^{N} x_{ij} \right)^2.
\end{equation}

For the sign constraints we can directly use the equality and difference penalties from the preliminaries:
\begin{equation}
    Q_{He}=\sum_{k=1}^{N_{He}}\left(x_{I_k^{He},J_k^{He}}-x_{I_k^{He},J_k^{He}+1}\right)^2,
\end{equation}
\begin{equation}
    Q_{Ve}=\sum_{k=1}^{N_{Ve}}\left(x_{I_k^{Ve},J_k^{Ve}}-x_{I_k^{Ve}+1,J_k^{Ve}}\right)^2,
\end{equation}
\begin{equation}
    Q_{Hc}=\sum_{k=1}^{N_{Hc}}\left(x_{I_k^{Hc},J_k^{Hc}}+x_{I_k^{Hc},J_k^{Hc}+1}-1\right)^2,
\end{equation}
\begin{equation}
    Q_{Vc}=\sum_{k=1}^{N_{Vc}}\left(x_{I_k^{Vc},J_k^{Vc}}+x_{I_k^{Vc}+1,J_k^{Vc}}-1\right)^2.
\end{equation}

The '=' and 'x' symbols are naturally described by a parity graph $G=(V,E)$ whose vertices are cells and whose edges carry a parity $p_e\in\{0,1\}$: $p_e=0$ for '=' and $p_e=1$ for 'x'. The constraint on an edge $e=(u,v)$ is
\begin{equation}
    x_u\oplus x_v = p_e.
\end{equation}

\begin{proposition}
The parity system on $G$ is consistent if and only if the XOR of the parities along every cycle of $G$ is $0$.
\end{proposition}

\begin{proof}
If one XORs the equations $x_u\oplus x_v=p_e$ along a cycle, every vertex variable appears twice and cancels, so any satisfying assignment must yield XOR $0$ for the edge parities. Conversely, if every cycle has XOR $0$, fixing one representative value in each connected component determines all other vertices uniquely by path parity, and the cycle condition guarantees that this value does not depend on the chosen path.
\end{proof}

Therefore the sign constraints can be reduced by a union-find structure with parity. If the system is inconsistent, the puzzle has no solution. If it is consistent, every connected component $C$ of the parity graph admits a representative variable $y_C$ and fixed parities $\pi_v\in\{0,1\}$ such that
\begin{equation}
    x_v = y_C \oplus \pi_v\quad \forall v\in C.
\end{equation}
If a clue fixes one vertex in the component, then it also fixes $y_C$; if two clues induce incompatible values, the puzzle has no solution. Consequently, the exact number of free variables after reducing the signs is the number of parity-connected components that are not fixed by clues. The expression
\[
    NM-N_{I0}-N_{I1}-N_{He}-N_{Hc}-N_{Ve}-N_{Vc}
\]
is only a crude upper bound when every sign contributes an independent reduction; in general it is not exact because cycles and redundant constraints do not reduce the number of free variables further.

At the level of a single edge, the previous reduction can still be written through the substitutions
\begin{itemize}
    \item In the case of having an '=' between $(i,j)$ and $(i,j+1)$
    \begin{equation}\label{eq: cambio 1}
        x_{i,j+1} = x_{i,j}.
    \end{equation}
    \item In the case of having an '=' between $(i,j)$ and $(i+1,j)$
    \begin{equation}\label{eq: cambio 2}
        x_{i+1,j} = x_{i,j}.
    \end{equation}
    \item In the case of having an 'x' between $(i,j)$ and $(i,j+1)$
    \begin{equation}\label{eq: cambio 3}
        x_{i,j+1} = (1-x_{i,j}).
    \end{equation}
    \item In the case of having an 'x' between $(i,j)$ and $(i+1,j)$
    \begin{equation}\label{eq: cambio 4}
        x_{i+1,j} = (1-x_{i,j}).
    \end{equation}
\end{itemize}
These local substitutions are single-edge instances of the parity-component reduction described above.

Hence the local TTP QUBO can be written as
\begin{equation}
    Q_{TTP}^{\mathrm{local}} =\ Q_{\mathrm{clues}} + Q_{HR} + Q_{VR} + Q_{Hr} + Q_{Vr}+ Q_{He}+Q_{Ve}+Q_{Hc}+Q_{Vc}.
\end{equation}
For Tango, or for the Takuzu/Tango variant without global row/column uniqueness, this formulation is exact after a consistent parity preprocessing. For classical Takuzu it is only a relaxation, because it does not yet enforce that the completed rows and columns be pairwise distinct. If one keeps the fractional forms $Q_{HR}$ and $Q_{VR}$, every feasible local board has the constant offset $(NM-N-M)/2$ coming from the length-3 windows; subtracting this constant does not change the minimizers.

\subsubsection{Exact Takuzu QUBO with witness variables}
We now restrict to the classical Takuzu board, which is square of size $N\times N$, and add an exact QUBO for the global non-repetition constraint without using slack variables.

For each pair of rows $r<s$ and each column $j$, introduce witness variables
\[
    u^{01}_{rsj},u^{10}_{rsj}\in\{0,1\}.
\]
The intended meaning is that $u^{01}_{rsj}=1$ certifies $x_{rj}=0$ and $x_{sj}=1$, while $u^{10}_{rsj}=1$ certifies $x_{rj}=1$ and $x_{sj}=0$. Define
\begin{equation}
    S^{\mathrm{row}}_{rs}=\sum_{j=1}^{N}\left(u^{01}_{rsj}+u^{10}_{rsj}\right).
\end{equation}
Then we add the penalty
\begin{equation}\label{eq: takuzu rows exact}
    Q^{\mathrm{row}}_{\neq}
    =
    A_{\neq}\sum_{1\leq r<s\leq N}
    \Bigg[
    \left(S^{\mathrm{row}}_{rs}-1\right)^2 +\sum_{j=1}^{N}
    \Big(
    u^{01}_{rsj} x_{rj}
    + u^{01}_{rsj}(1-x_{sj}) + u^{10}_{rsj}(1-x_{rj})
    + u^{10}_{rsj} x_{sj}
    \Big)
    \Bigg].
\end{equation}
The term $\left(S^{\mathrm{row}}_{rs}-1\right)^2$ forces exactly one active witness for each pair of rows, while the consistency terms force that an active witness must point to a column in which the two rows genuinely differ.

\begin{proposition}
For fixed board variables $x$, there exists an assignment of the row-witness variables $u$ with $Q^{\mathrm{row}}_{\neq}=0$ if and only if all rows are pairwise distinct.
\end{proposition}

\begin{proof}
Assume rows $r$ and $s$ are different. Then there exists a column $j$ with $x_{rj}\neq x_{sj}$. If $(x_{rj},x_{sj})=(0,1)$, set $u^{01}_{rsj}=1$; if $(x_{rj},x_{sj})=(1,0)$, set $u^{10}_{rsj}=1$; keep all other witnesses for that pair at $0$. Then $S^{\mathrm{row}}_{rs}=1$ and every consistency term vanishes. Conversely, if $Q^{\mathrm{row}}_{\neq}=0$, then for each pair $r<s$ exactly one witness is active. If it is $u^{01}_{rsj}$, the consistency terms force $x_{rj}=0$ and $x_{sj}=1$; if it is $u^{10}_{rsj}$, they force $x_{rj}=1$ and $x_{sj}=0$. In both cases the rows differ.
\end{proof}

Similarly, for each pair of columns $c<t$ and each row $i$, introduce
\[
    v^{01}_{cti},v^{10}_{cti}\in\{0,1\},
\]
where $v^{01}_{cti}=1$ certifies $x_{ic}=0$ and $x_{it}=1$, and $v^{10}_{cti}=1$ certifies $x_{ic}=1$ and $x_{it}=0$. Define
\begin{equation}
    S^{\mathrm{col}}_{ct}=\sum_{i=1}^{N}\left(v^{01}_{cti}+v^{10}_{cti}\right),
\end{equation}
and add
\begin{equation}\label{eq: takuzu cols exact}
    Q^{\mathrm{col}}_{\neq}
    =
    A_{\neq}\sum_{1\leq c<t\leq N}
    \Bigg[
    \left(S^{\mathrm{col}}_{ct}-1\right)^2 + \sum_{i=1}^{N}
    \Big(
    v^{01}_{cti} x_{ic}
    + v^{01}_{cti}(1-x_{it}) +  v^{10}_{cti}(1-x_{ic})
    + v^{10}_{cti} x_{it}
    \Big)
    \Bigg].
\end{equation}

\begin{proposition}
For fixed board variables $x$, there exists an assignment of the column-witness variables $v$ with $Q^{\mathrm{col}}_{\neq}=0$ if and only if all columns are pairwise distinct.
\end{proposition}

\begin{proof}
If columns $c$ and $t$ are different, choose one row $i$ in which they differ and activate the corresponding witness $v^{01}_{cti}$ or $v^{10}_{cti}$. Then $S^{\mathrm{col}}_{ct}=1$ and all consistency terms vanish. Conversely, if $Q^{\mathrm{col}}_{\neq}=0$, each pair $c<t$ has exactly one active witness, and that witness certifies a row in which the two columns differ.
\end{proof}

Let
\begin{equation}
    Q_{\mathrm{balance}}=Q_{Hr}+Q_{Vr},
\end{equation}
\begin{equation}
    Q_{\mathrm{triple}}=Q_{HR}+Q_{VR}.
\end{equation}
Then an exact QUBO for classical Takuzu is
\begin{equation}
    Q_{Takuzu}^{\mathrm{exact}}
    =
    Q_{\mathrm{clues}}
    + Q_{\mathrm{balance}}
    + Q_{\mathrm{triple}}
    + Q^{\mathrm{row}}_{\neq}
    + Q^{\mathrm{col}}_{\neq}.
\end{equation}
If one keeps the fractional form in $Q_{\mathrm{triple}}$, the feasible boards share a constant energy offset; the minimizing boards are nevertheless unchanged, and one may subtract the constant or use the zero-based triple penalty from the preliminaries if a zero feasible energy is preferred.

\begin{theorem}
Minimizing $Q_{Takuzu}^{\mathrm{exact}}$ over the board variables and the witness variables produces exactly the solutions of the full Takuzu problem.
\end{theorem}

\begin{proof}
The term $Q_{\mathrm{clues}}$ enforces the clues, $Q_{\mathrm{balance}}$ enforces the row and column balance conditions, and $Q_{\mathrm{triple}}$ enforces that no horizontal or vertical length-3 window is all-zero or all-one. By the two previous propositions, $Q^{\mathrm{row}}_{\neq}$ enforces that no two rows coincide and $Q^{\mathrm{col}}_{\neq}$ enforces that no two columns coincide. Therefore every minimizer satisfies exactly the Takuzu rules. Conversely, if a board is a valid Takuzu solution, then every pair of rows and every pair of columns differ somewhere, so one can activate one consistent witness for each pair and obtain zero contribution from the new uniqueness terms; hence valid Takuzu boards are precisely the minimizing configurations.
\end{proof}

The row witnesses use $2N\binom{N}{2}=N^2(N-1)$ auxiliary variables, and the column witnesses use the same amount. Therefore the exact Takuzu construction requires
\[
    2N^2(N-1)=O(N^3)
\]
auxiliary binary variables in total, and no slack variables are needed for the row/column uniqueness constraint. This shows that the global uniqueness condition can indeed be incorporated into an exact QUBO formulation, at the cost of auxiliary witness variables.

\subsection{Generalizations of the TTP}
Having solved the original problem, we will present several generalizations of TTP.

\subsubsection{Unbalanced regularity condition}
The first generalization we propose is to allow for each row or column to have a different number of zeros than ones. Since we know that the number of ones determines the number of zeros, we can say that if there are $N$ cells in that direction, we want there to be $N_1$ ones and $N-N_1$ zeros. This can also change between rows and columns. We say that in row $i$ we want $M_{1,i}$ ones, and in column $j$ we want $N_{1,j}$ ones. These generalized terms, for both horizontal and vertical, are
\begin{equation}
    Q_{Hr} = \sum_{i=1}^{N}\left(M_{1,i} - \sum_{j=1}^{N} x_{ij} \right)^2,
\end{equation}
\begin{equation}
    Q_{Vr} = \sum_{j=1}^{M}\left(N_{1,j} - \sum_{i=1}^{N} x_{ij} \right)^2.
\end{equation}

\subsubsection{Diagonal repetition}
The second generalization we propose is that, in addition to not being able to have more than two ones or zeros in a row vertically or horizontally, it should not be possible to have more than two ones or zeros diagonally. For this, we define 2 new terms, exactly the same as the original ones for this, but changing the second summation, so that they evaluate the lower diagonals for each cell. We have $Q_{DLR}$ for the left diagonals and $Q_{DRR}$ for the right diagonals:
\begin{equation}
    Q_{DLR} = \sum_{i,j=1,3}^{N-2,M}\left(\frac{3}{2} - \sum_{k=0}^{2} x_{i+k,j-k} \right)^2,
\end{equation}
\begin{equation}
    Q_{DRR} = \sum_{i,j=1}^{N-2,M-2}\left(\frac{3}{2} - \sum_{k=0}^{2} x_{i+k,j+k} \right)^2,
\end{equation}
being $x_{a,b}= 0 \quad \forall b<1, a\in [1,N]$, and $\forall b>M, a\in [0,N-1]$.

\subsubsection{Regions condition}
\begin{figure}
    \centering
    \includegraphics[width=\linewidth]{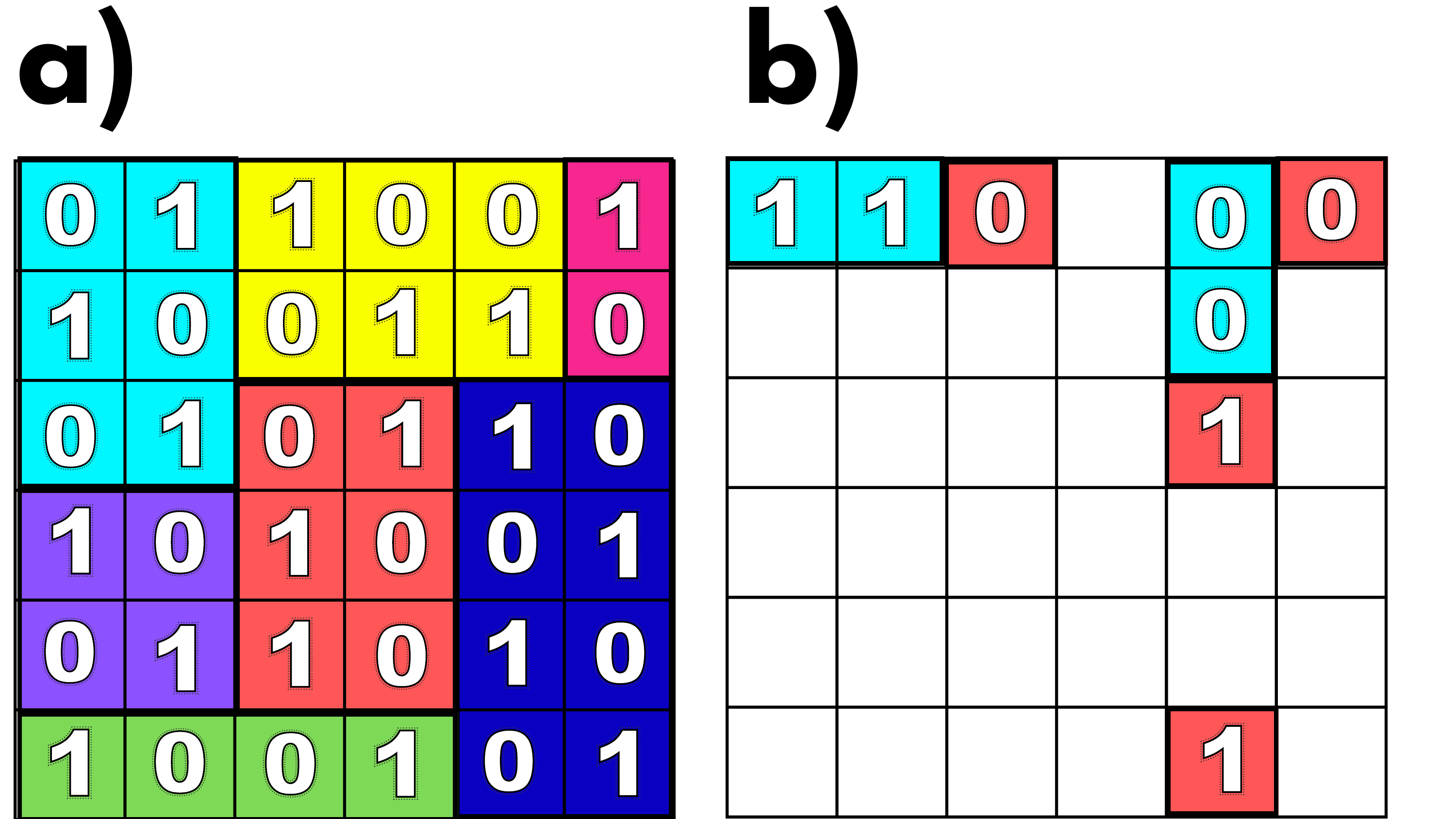}
    \caption{Generalized cases. a) Takuzu with colored regions, b) toroidal Takuzu, where the blue cells have fixed initial values and the red cells are a consequence of the repetition constraint.}
    \label{fig: Generals}
\end{figure}

Another possible generalization is the imposition of a specific number of ones in certain colored regions. That is, for a set of $N_R$ regions, in the $k$-th region $R_k$ there must be $m_k$ ones, as we can see in Fig. \ref{fig: Generals} a. This condition is imposed with a term
\begin{equation}
    \sum_{k=1}^{N_R} \left(m_k - \sum_{(i,j) \in R_k} x_{ij}\right)^2.
\end{equation}

\subsubsection{Toroidal board}
An interesting generalization is to consider a toroidal board, so that the condition that there are no more than 2 ones or zeros in a row vertically or horizontally is cyclically extended to the other end of the board, as can be seen in Fig. \ref{fig: Generals} b.

The only term that will change will be the repetition constraint, so that we will have to perform the summations with all rows and columns, making $x_{a,b}=x_{a-N,b}$ if $a>N$ and $x_{a,b}=x_{a,b-M}$ if $b>M$. In this way we will take into account the toroid shape.  In case we consider a cylinder instead of a toroid, we will only have to keep one of these changes, in the direction that is cyclic.

\subsubsection{Long distance equal/cross}
\begin{figure}
    \centering
    \includegraphics[width=\linewidth]{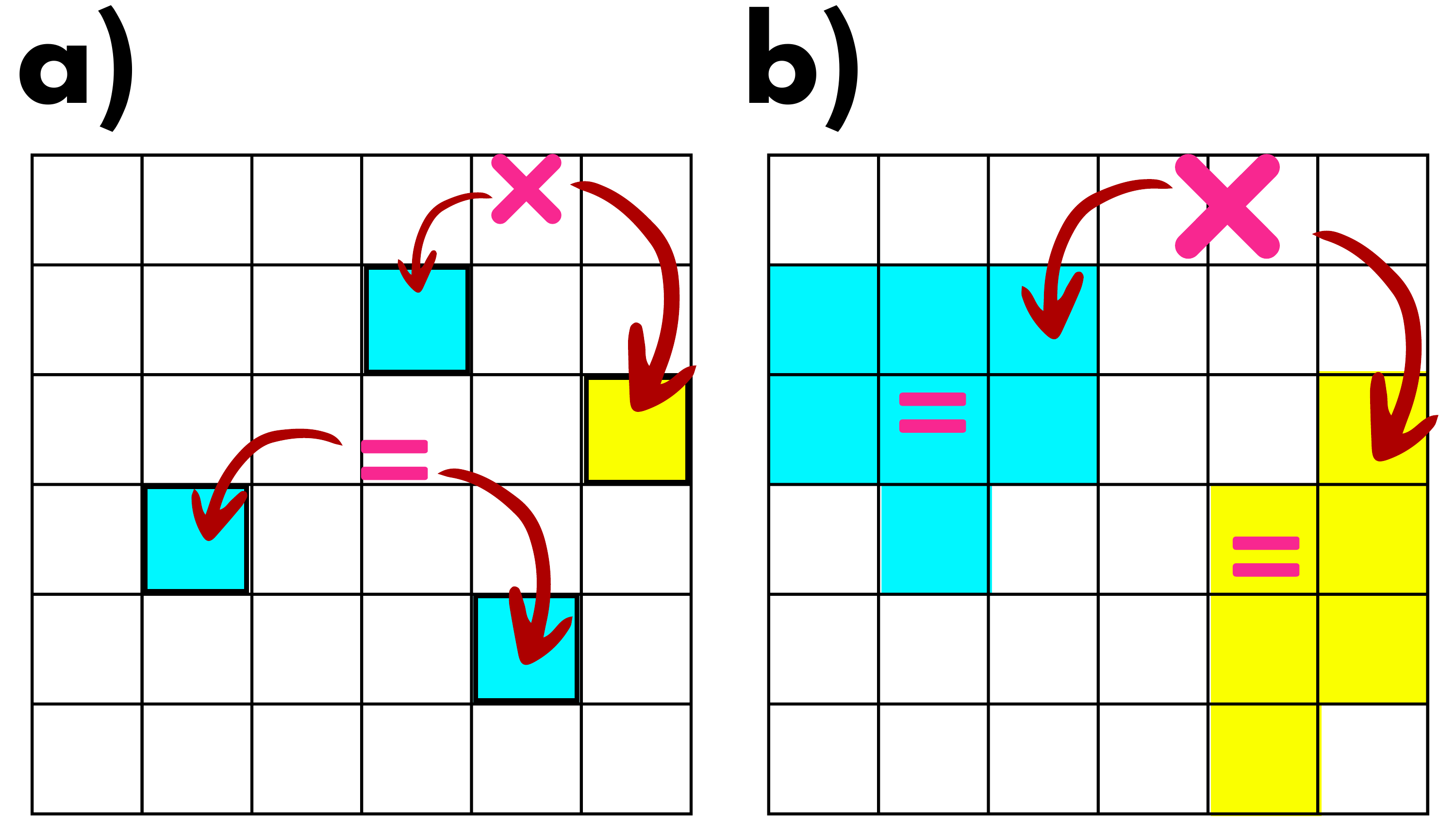}
    \caption{a) Symbols that act over long distances. b) Regions of equality that are different from each other. All cells in the blue region have the same value, all cells in the yellow region have the same value, and the blue cells all have different values than the yellow cells.}
    \label{fig: Long_distance}
\end{figure}

In Tango's original problem, we have that there can only be '=' and 'x' symbols acting between vertically and horizontally adjacent cells. We extend this idea to mean that both symbols can act between any two cells, as can be seen in Fig. \ref{fig: Long_distance} a. To implement this generalization, we need only allow that the variable changes exhibited in \eqref{eq: cambio 1}, \eqref{eq: cambio 2}, \eqref{eq: cambio 3} and \eqref{eq: cambio 4} can be made between two non-adjacent variables. The rest of the formulation is the same.

On the other hand, we also generalize the possibility of defining regions in which all their variables have to be equal (equality regions), and separately defining that a pair of equality regions have to have different values for their variables from each other, as in Fig. \ref{fig: Long_distance} b. To do this, just impose a set of '=' symbols between the cells of the region, so that they form a chain of equalities with each other, and then impose a single 'x' symbol between any pair of cells of one and the other region. In this way, since all cells in each region have to be equal, if a cell in one is different from a cell in the other, all cells in one region will be opposite cells in the other region.

\paragraph{Summary of exactness.}
The N-Queens formulation is exact when one combines row and column equalities with the diagonal conflict graph. The LinkedIn Queens formulation is exact after preprocessing the pre-placed queens. The local TTP formulation is exact for Tango, after a consistent parity reduction, and for Takuzu/Tango variants without global row/column uniqueness, but it is only a relaxation of classical Takuzu. The witness-variable construction yields an exact Takuzu QUBO with $2N^2(N-1)=O(N^3)$ auxiliary variables and no slack variables for the uniqueness constraint. For Tents \& Trees, the neighborhood-only version is a relaxation, whereas the formulation with assignment variables $y_{t,c}$ is exact. The Max Chess Pieces formulation is exact provided that $\lambda>\max_v w_v$, and in the unweighted case it is enough to take $\lambda>1$.

\section{Conclusions}
We have presented several generalizations of the N-queens problem, including the LinkedIn version, and clarified that their diagonal restrictions are most naturally expressed through pairwise conflict graphs rather than through arbitrary unions of diagonals treated as single regions. This preserves the exactness of the N-queens and LinkedIn Queens formulations and makes the preprocessing of pre-placed queens fully rigorous, including the need to remove adjacent-diagonal conflicts in the LinkedIn case. We have also revisited Tents \& Trees and the chess-piece variants, distinguishing carefully between simplified relaxations and exact formulations.

For Takuzu and Tango, the local QUBO captures the clues, the balance constraints, the prohibition of three equal consecutive values and the parity signs. This local formulation is exact for Tango, after a consistent parity reduction, and for Takuzu/Tango variants without global row/column uniqueness. For classical Takuzu, however, the local model is only a relaxation. We have shown that the missing global uniqueness condition can be incorporated exactly by means of row and column witness variables, obtaining an exact Takuzu QUBO at the cost of $2N^2(N-1)=O(N^3)$ auxiliary binary variables and without using slack variables for that purpose. We have also kept the fractional terms centered at $q+\frac{1}{2}$ as a useful slack-saving technique, explicitly noting that they introduce only a constant feasible-energy offset and therefore do not affect the minimizing assignments.

Future lines of research may include its implementation in various types of quantum hardware, benchmarking these families of puzzles, reducing the auxiliary-variable overhead of exact global uniqueness constraints, extending the region penalties to longer admissible occupancy intervals, creating a Grover's \cite{Grover} oracle to solve the problem, and studying related QUDO or HOBO formulations.

\section*{Acknowledgment}
This work is framed within the Quipucamayocs, the activities of the QuantumQuipu community.

\bibliographystyle{unsrt}  
\bibliography{references}

@PREAMBLE{
 "\providecommand{\noopsort}[1]{}" 
 # "\providecommand{\singleletter}[1]{#1}%" 
}

@misc{QAOA,
      title={A Quantum Approximate Optimization Algorithm}, 
      author={Edward Farhi and Jeffrey Goldstone and Sam Gutmann},
      year={2014},
      eprint={1411.4028},
      archivePrefix={arXiv},
      primaryClass={quant-ph}
}

@article{Annealing_Industry,
   title={Quantum annealing for industry applications: introduction and review},
   volume={85},
   ISSN={1361-6633},
   url={http://dx.doi.org/10.1088/1361-6633/ac8c54},
   DOI={10.1088/1361-6633/ac8c54},
   number={10},
   journal={Reports on Progress in Physics},
   publisher={IOP Publishing},
   author={Yarkoni, Sheir and Raponi, Elena and Bäck, Thomas and Schmitt, Sebastian},
   year={2022},
   month=sep, pages={104001} }

@article{Annealing_Overview,
   title={Quantum annealing: an overview},
   volume={381},
   ISSN={1471-2962},
   url={http://dx.doi.org/10.1098/rsta.2021.0417},
   DOI={10.1098/rsta.2021.0417},
   number={2241},
   journal={Philosophical Transactions of the Royal Society A: Mathematical, Physical and Engineering Sciences},
   publisher={The Royal Society},
   author={Rajak, Atanu and Suzuki, Sei and Dutta, Amit and Chakrabarti, Bikas K.},
   year={2022},
   month=dec }

@misc{QUBO,
      title={A Tutorial on Formulating and Using QUBO Models}, 
      author={Fred Glover and Gary Kochenberger and Yu Du},
      year={2019},
      eprint={1811.11538},
      archivePrefix={arXiv},
      primaryClass={cs.DS}
}

@misc{N_Queens,
      title={The $n$-queens problem}, 
      author={Candida Bowtell and Peter Keevash},
      year={2021},
      eprint={2109.08083},
      archivePrefix={arXiv},
      primaryClass={math.CO},
      url={https://arxiv.org/abs/2109.08083}, 
}

@article{Quantum_N_Queens,
   title={A Quantum N-Queens Solver},
   volume={3},
   ISSN={2521-327X},
   url={http://dx.doi.org/10.22331/q-2019-06-03-149},
   DOI={10.22331/q-2019-06-03-149},
   journal={Quantum},
   publisher={Verein zur Forderung des Open Access Publizierens in den Quantenwissenschaften},
   author={Torggler, Valentin and Aumann, Philipp and Ritsch, Helmut and Lechner, Wolfgang},
   year={2019},
   month=jun, pages={149} }

@inproceedings{Sudoku_QUBO,
author = {M\"{u}cke, Sascha},
title = {A Simple QUBO Formulation of Sudoku},
year = {2024},
isbn = {9798400704956},
publisher = {Association for Computing Machinery},
address = {New York, NY, USA},
url = {https://doi.org/10.1145/3638530.3664106},
doi = {10.1145/3638530.3664106},
booktitle = {Proceedings of the Genetic and Evolutionary Computation Conference Companion},
pages = {1958–1962},
numpages = {5},
location = {Melbourne, VIC, Australia},
series = {GECCO '24 Companion}
}

@misc{Fractionary,
      title={Qubit Number Optimization for Restriction Terms of QUBO Hamiltonians}, 
      author={Iñigo Perez Delgado and Beatriz García Markaida and Alejandro Mata Ali and Aitor Moreno Fdez. de Leceta},
      year={2023},
      eprint={2306.06943},
      archivePrefix={arXiv},
      primaryClass={quant-ph},
      url={https://arxiv.org/abs/2306.06943}, 
}

@INPROCEEDINGS{Old_QUBO_N_queens,
  author={Tsukiyama, Shunsuke and Nakano, Koji and Ito, Yasuaki and Yazane, Takashi and Yano, Junko and Kato, Takumi and Ozaki, Shiro and Mori, Rie and Katsuki, Ryota},
  booktitle={2023 Eleventh International Symposium on Computing and Networking (CANDAR)}, 
  title={Solving the N-Queens Puzzle by a QUBO Model with Quadratic Size}, 
  year={2023},
  volume={},
  number={},
  pages={59-67},
  doi={10.1109/CANDAR60563.2023.00015}}

@inproceedings{At_most_QUBO_queens,
author = {Codognet, Philippe},
title = {Encoding the At-Most-One Constraint for QUBO and Quantum Annealing: Experiments with the N-Queens problem},
year = {2023},
isbn = {9798400701207},
publisher = {Association for Computing Machinery},
address = {New York, NY, USA},
url = {https://doi.org/10.1145/3583133.3596394},
doi = {10.1145/3583133.3596394},
booktitle = {Proceedings of the Companion Conference on Genetic and Evolutionary Computation},
pages = {2195–2202},
numpages = {8},
location = {Lisbon, Portugal},
series = {GECCO '23 Companion}
}

@misc{Quantum_Alg_Queens,
      title={A Quantum Approach to solve N-Queens Problem}, 
      author={Santhosh G S and Piyush Joshi and Ayan Barui and Prasanta K. Panigrahi},
      year={2023},
      eprint={2312.16312},
      archivePrefix={arXiv},
      primaryClass={quant-ph},
      url={https://arxiv.org/abs/2312.16312}, 
}

@article{Binary_Genetic,
title = {Solving the Binary Puzzle with Genetic Algorithm},
journal = {Procedia Computer Science},
volume = {234},
pages = {954-961},
year = {2024},
note = {Seventh Information Systems International Conference (ISICO 2023)},
issn = {1877-0509},
doi = {https://doi.org/10.1016/j.procs.2024.03.084},
url = {https://www.sciencedirect.com/science/article/pii/S1877050924004423},
author = {Rachel Anne B. Balagbis and Orven E. Llantos}
}

@InProceedings{Takuzu_quantum,
author="Foug{\`e}res, Alain-J{\'e}r{\^o}me",
editor="Nguyen, Ngoc Thanh
and Papadopoulos, George A.
and J{\k{e}}drzejowicz, Piotr
and Trawi{\'{n}}ski, Bogdan
and Vossen, Gottfried",
title="Agent Having Quantum Properties: The Superposition States and the Entanglement",
booktitle="Computational Collective Intelligence",
year="2017",
publisher="Springer International Publishing",
address="Cham",
pages="389--398",
isbn="978-3-319-67074-4"
}

@inproceedings{Grover,
author = {Grover, Lov K.},
title = {A fast quantum mechanical algorithm for database search},
year = {1996},
isbn = {0897917855},
publisher = {Association for Computing Machinery},
address = {New York, NY, USA},
url = {https://doi.org/10.1145/237814.237866},
doi = {10.1145/237814.237866},
booktitle = {Proceedings of the Twenty-Eighth Annual ACM Symposium on Theory of Computing},
pages = {212–219},
numpages = {8},
location = {Philadelphia, Pennsylvania, USA},
series = {STOC '96}
}

\end{document}